\PassOptionsToPackage{pdftex,final,breaklinks,hidelinks,colorlinks=false,bookmarks=false,bookmarksnumbered,plainpages=false,pdfpagelabels,pdfborder={0 0 0},hyperindex}{hyperref}
\documentclass[manuscript]{acmart} 
\usepackage{array}
\usepackage{stfloats}
\usepackage{url}
\usepackage{verbatim}
\usepackage[table,svgnames]{xcolor}
\usepackage{color}
\usepackage{bm}
\usepackage{multirow} 
\usepackage{listings}
\usepackage{enumitem}
\usepackage{tikz}
\usetikzlibrary{calc, positioning, matrix, shapes, shapes.callouts, shapes.misc, shapes.symbols, shapes.geometric, shapes.multipart, fit, shadows, decorations, shapes.arrows, arrows, trees, mindmap, intersections, backgrounds, positioning, decorations.shapes, decorations.pathreplacing, decorations.pathmorphing, decorations.markings}
\tikzstyle{every picture}+=[remember picture]
\tikzstyle{na} = [baseline=-2.5pt]  
\tikzstyle{block} = [rectangle, draw, fill=white!20,
text width=12em, text centered, rounded corners, minimum height=4em]
\tikzstyle{line} = [draw, -latex']
\usepackage[noteubner]{showcode}

\usepackage{makecell}
\usepackage{soul}

\usepackage{graphicx}
\usepackage{subcaption}

\definecolor{color1}{RGB}{192, 188, 181}
\definecolor{color2}{RGB}{74, 108, 111}
\definecolor{color3}{RGB}{132, 96, 117}
\definecolor{color4}{RGB}{175, 93, 99}
\definecolor{color5}{RGB}{237, 71, 74}

\definecolor{codegreen}{RGB}{56,118,29}
\definecolor{backcolour}{RGB}{255,242,204}
\definecolor{codered}{RGB}{224,104,103}
\definecolor{codeblue}{RGB}{60,120,216}

\tcbset{
    center title,
    left=2pt,
    right=2pt,
    top=2pt,
    bottom=2pt,
    colback=gray!10,
    colframe=gray!60
}

\AtBeginDocument{%
  }

\usepackage{array}

\begin{document}

\clearpage
\thispagestyle{empty}

\begin{center}
\large\bfseries
This work has been submitted to ACM for possible publication.\\
Copyright may be transferred without notice, after which this version may no longer be accessible.
\end{center}

\clearpage

\title{Comprehension vs. Adoption: Evaluating a Language Workbench Through a Family of Experiments}

\author{Giovanna Broccia}
\authornote{Corresponding Authors.}
\email{giovanna.broccia@isti.cnr.it}
\orcid{0000-0002-4737-5761}
\affiliation{
  \institution{CNR–ISTI}
  \city{Pisa}
  \country{Italy}
}

\author{Maurice H. ter Beek}
\email{maurice.terbeek@isti.cnr.it}
\affiliation{
  \institution{CNR–ISTI}
  \city{Pisa}
  \country{Italy}
}

\author{Walter Cazzola}
\affiliation{%
  \institution{University of Milan}
  \city{Milan}
  \country{Italy}}
\email{cazzola@di.unimi.it}

\author{Luca Favalli}
\affiliation{%
  \institution{University of Milan}
  \city{Milan}
  \country{Italy}}
\email{luca.favalli@unimi.it}

\author{Francesco Bertolotti}
\affiliation{%
  \institution{University of Milan}
  \city{Milan}
  \country{Italy}}

\author{Alessio Ferrari}
\authornotemark[1]
\affiliation{%
  \institution{University College Dublin}
  \city{Dublin}
  \country{Ireland}}
\affiliation{%
  \institution{CNR-ISTI}
  \city{Pisa}
  \country{Italy}
}
\email{alessio.ferrari@ucd.ie}

\renewcommand{\shortauthors}{Broccia et al.}

\begin{abstract}
Language workbenches are tools that enable the definition, reuse, and composition of programming languages and their ecosystems, aiming to streamline language development. To facilitate their adoption by language designers, the \textit{comprehensibility} of the meta-language used in these workbenches (i.e., the language used to define other languages) is an important aspect to consider and evaluate. Moreover, considering that language workbenches are relatively new tools and still not widely spread, \textit{user acceptance} emerges as a crucial factor to be accounted for during their assessment. Current literature often neglects user-centred aspects like comprehensibility and acceptance in the assessment of this breed of tools. This paper addresses this gap through a family of experiments assessing Neverlang, a modular language workbench. 
The study adopts a tailored version of the Method Evaluation Model (MEM) to evaluate the comprehensibility of Neverlang’s meta-language and programs, as well as user acceptance in terms of perceived ease of use, perceived usefulness, and intention to use. It also investigates the relationships among these dimensions. The experiments were conducted in three iterations—an original study followed by two replications—involving participants from academia.
The results reveal that users demonstrate sufficient comprehension of Neverlang's meta-language, particularly concerning its syntax, express a favourable perception of its usefulness, and indicate their intention to use it. However, the results also indicate that Neverlang’s ease of use remains a challenge. 
Additionally, variations in the perceived ease of use, and perceived usefulness---whether low or high---influence the users' intention to use the tool.
Surprisingly, no significant correlation is found between comprehensibility and user acceptance. 
These findings highlight the need to consider both technical and human-centric factors when evaluating language workbenches. Notably, higher comprehensibility of the meta-language does not necessarily translate into greater acceptance, underscoring the complex interplay between comprehension %usability?
and adoption.

\end{abstract}

\begin{CCSXML}
<ccs2012>
   <concept>
       <concept_id>10011007.10011006.10011008</concept_id>
       <concept_desc>Software and its engineering~General programming languages</concept_desc>
       <concept_significance>500</concept_significance>
       </concept>
   <concept>
       <concept_id>10011007.10011006.10011066</concept_id>
       <concept_desc>Software and its engineering~Development frameworks and environments</concept_desc>
       <concept_significance>500</concept_significance>
       </concept>
   <concept>
       <concept_id>10011007.10011006.10011050.10011017</concept_id>
       <concept_desc>Software and its engineering~Domain specific languages</concept_desc>
       <concept_significance>500</concept_significance>
       </concept>
 </ccs2012>
\end{CCSXML}

\ccsdesc[500]{Software and its engineering~General programming languages}
\ccsdesc[500]{Software and its engineering~Development frameworks and environments}
\ccsdesc[500]{Software and its engineering~Domain specific languages}

\keywords{family of experiments, quasi-experiment, language workbenches, comprehensibility, user acceptance, MEM, Neverlang, empirical software engineering}

\maketitle

\section{Introduction}
Language workbenches~\cite{Fowler05} are tools that support the development of 
general-purpose and domain-specific 
programming languages, their implementation, and the generation of their ecosystem, including but not limited to a dedicated integrated development environment (IDE). This breed of tools exploits the concepts of \textit{language feature} and \textit{component}~\cite{Cazzola17c} to modularise the specification and implementation of the language under development. 
Such modularisation makes the creation of new languages easier and more affordable by reusing existing language components~\cite{erdweg2013state}. 

Most language workbenches are academic products that must be evaluated for comprehensibility, before being released to a wider public.
Existing research has predominantly focused on assessing the complexity of language development tasks in terms of quantitative aspects such as lines of code, amount of external dependencies (both static and dynamic), and time needed to complete the task~\cite{erdweg2013state, Erdweg15, kelly2013empirical}.
However, there remains a notable gap in the literature concerning user-centred aspects, such as the comprehensibility of the meta-language used in these workbenches (i.e., a specialised language for defining other languages), and users' perceptions and acceptance of these tools.

Language \textit{comprehension} stands as a cornerstone in the utilisation of (new) languages and tools, significantly influencing user performance, and it has been a well-studied field for over 40 years~\cite{WyrichBW24}. 
The need for comprehensible programming languages is well recognised in software development, where developers spend between 50\% and 70\% of their daily work time on understanding code~\cite{minelli2015know, xia2017measuring}. 
Although these studies primarily examine general-purpose languages (GPLs) like Smalltalk, Java, and C\#, we contend that understanding meta-languages is an equally central aspect of working with language workbenches. 

Language comprehension is often measured through performance-based variables, such as the effectiveness of users in performing certain tasks using the language under analysis~\cite{abrahao2011evaluating, kosar2012program}. The \textit{user acceptance} of new technologies and notations further impacts their actual adoption and is typically assessed through perception-based variables, such as ease of use, usefulness, and intention to use~\cite{venkatesh2003user}.  Studies also suggest a correlation between perception-based variables and user comprehension of certain notations measured through performance-based variables, indicating that comprehension is often associated with user acceptance~\cite{abrahao2011evaluating, davis1989perceived}. 

\textbf{Study Design.} This paper presents a family of three experiments~\cite{basili1999building} aiming at evaluating the comprehensibility and users' acceptance of 
a language workbench named Neverlang~\cite{Cazzola15c}.  
We select Neverlang as a representative case, as it shares most of the standard features of the other workbenches, such as satisfying all the constraints of the \textit{language extension problem}~\cite{Leduc20} and all the language composition mechanisms defined by Erdweg et al.~\cite{Erdweg12}.
Moreover, language components written in Neverlang have already been evaluated through software-related metrics such as cohesion, coupling, cognitive complexity, and maintainability index~\cite{CF22}. 
Since Neverlang is a novel tool that has not yet seen widespread adoption, we focus on novice users—those with no prior experience with the tool.

The original experiment was conducted in 2022, involving computer science undergraduate, master, and PhD students from the University of Milan---partial results outlined in~\cite{broccia2023evaluating}. To bolster the validity of the original experiment's findings, two differentiated independent replications were conducted. The first replication occurred in 2023 with subjects from the Bertinoro International Spring School (BISS), a doctoral school offering postgraduate-level courses for PhD students across Europe and beyond. The second replication took place in 2023 with computer science undergraduate, master, and PhD students from the University of Milan. These experiments will be referred to as ``UniMI~1 2022,'' ``BISS 2023,'' and ``UniMI~2 2023,'' respectively.

To evaluate Neverlang, we adapted the Method Evaluation Model (MEM)~\cite{moody2001dealing},
a model used to evaluate new information technologies, tools, and notations. 
It provides mechanisms for evaluating both the likelihood of acceptance and the actual impact of a method in practice.
MEM has been applied in the field of software engineering to forecast the understandability of requirement models~\cite{abrahao2011evaluating} and to gauge users' acceptance of attack-defense trees for modelling security requirements~\cite{broccia2024assessing}. In our analysis, we tailored MEM to evaluate and relate user comprehension and user acceptance of Neverlang and its meta-language.  Comprehension was evaluated through a test in which participants were required to perform a set of tasks related to syntax, semantics, and usage of the Neverlang meta-language. User acceptance was measured through a questionnaire, evaluating perceived ease of use, perceived usefulness, and intention to use.

We conducted hypothesis testing to examine the relationships among these variables. Furthermore, to aggregate and interpret results from our family of experiments, we carried out two types of meta-analyses: a meta-analysis of proportions focusing on key factors such as comprehensibility and user acceptance, and a meta-analysis of correlation coefficients to synthesise the observed relationships among acceptance and comprehensibility variables.

Although comprehensibility and user acceptance are conceptually related to usability, this study does not constitute a classical usability evaluation focused on interaction design aspects~\cite{preece2001interaction}. Instead, it investigates how novice users---after a structured training session---understand the meta-language of a language workbench and how they perceive its ease of use, usefulness, and their intention to use it. The evaluation reflects an educational perspective on comprehensibility and conceptual learning, rather than an assessment of user interaction with the tool.

\textbf{Results.} 
%With some differences, the results of the three experiments are consistent. Novice users demonstrate sufficient understanding of the Neverlang meta-language, and express positive opinions toward its usefulness and intention to use. No notable correlation is observed between Neverlang's comprehensibility and user acceptance. Ease of use is also lower than the other perception-based variables, indicating that the user-friendliness of the language is not its main strength. Nonetheless, users' intention to use Neverlang is correlated with both its perceived usefulness and 
%its ease of use,  
The outcomes of the data analysis demonstrate a consistent and statistically significant proportion of users with a satisfactory understanding of Neverlang across the individual studies: nearly 80\% of users across the included studies attained comprehensibility above the deemed sufficient threshold. 
Regarding user acceptance, the meta-analysis of proportions confirms that perceived usefulness is the only acceptance factor rated significantly above the neutral threshold, with 92\% of users recognising the tool as useful. In contrast, intention to use (62\%) and ease of use (48\%) did not reach statistical significance, suggesting that usability remains a challenge and that user willingness to adopt the tool is more uncertain. 
The meta-analysis of correlation coefficients provides additional insights: it indicates that the intention to use Neverlang is significantly associated with both perceived usefulness and perceived ease of use, suggesting that variations in these aspects---whether high or low---can influence users' willingness to adopt or not the tool.

The study suggests that Neverlang has potential as a tool for modular and declarative programming but needs to be more accessible and user-friendly. This includes developing comprehensive libraries, detailed documentation, and long-term, incremental learning programs. Balancing its powerful features with usability is essential, though resource limitations and the academic nature of language workbenches may hinder these improvements.

The design of this study targeting language workbenches establishes an analysis framework that can be used to 
evaluate other language workbenches' meta-languages. This can foster a better understanding of these platforms that are currently not widely studied, particularly from user-centred perspectives.

We make our replication package, including the material used in the experiments, their results, and the code for the statistical tests publicly  available~\cite{broccia_2025_16094518}. 

\textbf{Structure.} The remainder of this paper is structured as follows. Section~\ref{sect:background} provides background information on Neverlang.  Section~\ref{sec:studyDesign} outlines the experimental design, while Section~\ref{sect:individualExp} details the individual experiments.
The data analysis procedures and results from both the individual experiments and the family of experiments are presented in Section~\ref{sec:results}, while Section~\ref{sect:discussion} presents the discussion of the results. Sections~\ref{sect:threats} and~\ref{sect:related} address the threats to validity and the related work, respectively. Finally, Section~\ref{sect:conclusion} summarises the conclusions and suggests directions for future research. 

\section{Background}\label{sect:background}
This section provides the necessary background on the fundamental elements of language workbenches required to understand the remainder of the paper.

\textbf{Domain-Specific Languages.}
A \textit{domain-specific language} (DSL) is a programming language with limited expressiveness, tailored to express solutions in their target domain~\cite{VanDeursen00, Mernik05}.
DSLs are categorized as \textit{internal} (or embedded~\cite{Fowler10}) and \textit{external}. Internal DSLs are built within an existing \textit{host language}, leveraging its features; languages such as \textsf{Lisp}, \textsf{Ruby}, and \textsf{Scala} are well-suited for internal DSLs. Internal DSLs are usually considered easy to develop, since they do not require a dedicated parser and custom tooling and can instead leverage those offered those offered by the host language, but they are limited in the sense that they are implemented as \textit{fluent APIs}~\cite{Fowler05b} and must therefore adhere to the host syntax, limiting the possibility to fully express the target domain.
Conversely, external DSLs are standalone languages, requiring a full design of syntax and semantics. They offer greater freedom in notation and are often more accessible to domain experts, as any abstraction can be mapped to a language construct---or a keyword---along with its semantics.
However, external DSLs are harder to implement, as al tools---including parsers, optimizers, translators, editors, debuggers, etc.---must be implemented from scratch.
Notable external DSLs include \textsf{SQL} (database querying), \textsf{LaTeX} (document typesetting), and \textsf{Make} (build automation).

\textbf{Language-Oriented Programming.}
The \textit{language-oriented programming} paradigm~\cite{Ward94}, suggests that complex software systems can be built around a set of domain-specific languages to properly express domain problems and their solutions.
Although early definitions emphasized projectional editing~\cite{Fowler05}, the notion can be generalized to any attempt at incorporating several DSLs into a unique system, by supporting efficient language definition, reuse, and composition.
Typically, abstractions in libraries at subsystem level are delivered by means of design patterns~\cite{Gamma95}, which allow for standardized and reproducible solutions. However, the implementation linked to these patterns is often non-reusable due to application-specific quirks and configurations. A suitable solution is to use a very small programming language to implement the subsystem or the configuration itself. With regards to configurations, this is actually a very common practice and involves languages such as \textsf{XML}, \textsf{JSON}, \textsf{TOML}, and so on, and it dates back to the Unix tradition of \textit{little languages}~\cite{Bentley86}.
Opting for a language-oriented design allows non-programmers and domain-experts to be involved in the development activities, improving the communication among team members.
Despite its advantages, language-oriented programming never took center stage among mainstream programming paradigms---except for domains such as persistence (\textsf{SQL}), and the aforementioned configurations. This is mainly due to the complexity related to developing programming languages and their ecosystem: in most common cases, developing a module is considered more feasible than developing a full programming language and any related productivity tools.

\textbf{Language Workbenches.}
The term ``language workbench'' was introduced by Martin Fowler~\cite{Fowler05}, to describe tools suitable for supporting the language-oriented programming paradigm~\cite{Ward94}. Language workbenches provide custom DSLs for specifying and implementing syntax and semantics, as well as sophisticated composition techniques, but are still rarely used in practice~\cite{Ozkaya21}.
Language workbenches are varied in their nature, but we can broadly classify them in two groups: modelware and grammarware. Modelware language workbenches employ modelling techniques to define DSLs, which are based on metamodels and model transformations. Prominent examples are the tools based on the \textsf{Eclipse Modeling Framework} (EMF)~\cite{Steinberg08}, and the popular \textsf{JetBrains MPS}~\cite{Volter11}. Conversely, grammarware language workbenches are more akin to traditional programming languages, since they rely on textual grammars such as EBNF\@. Some examples are \textsf{Spoofax}~\cite{Wachsmuth14} and \textsf{Racket}~\cite{Flatt11}.
Many recent language workbenches~\cite{Erdweg15}---including \textsf{Spoofax}, \textsf{MPS}, \textsf{MontiCore}~\cite{Krahn10}, \textsf{LISA}~\cite{Henriques05}, and \textsf{Melange}~\cite{Degueule15}---tackle the limitations to th applicability of the language-oriented programming by offering tools for system designers, supporting IDEs generation via templates and the language server protocol, even though often reusability of collateral tools is limited.
One approch to ease DSL development is through composition mechanisms: \textsf{MontiCore} promotes compositional development of language families through mechanisms like language embedding and inheritance. Butting~\emph{et~al.}~\cite{Butting18} introduced a method to manage syntactic variability in extensible LPLs using \textsf{MontiCore}. \textsf{Spoofax} facilitates the generation of a range of IDE tools for both \textsf{Eclipse} and \textsf{IntelliJ}, including syntax highlighting, code completion, and parse error recovery, as well as language configuration capabilities.
Among existing tools, \textsf{MPS} arguably represents the most complete environment for developing language families, as it provides comprehensive IDE support and enables extensive customization of abstract syntax trees.
Nevertheless, this work focuses on the Neverlang language workbench mainly due to three factors:
\begin{enumerate}
    \item it provides a fine-grained and exogenous approach to language composition~\cite{Cazzola23b} that allows all its artifacts to be reused across a variety of unrelated DSLs with very little boilerplate code;
    \item it supports the feature-oriented paradigm~\cite{Prehofer97} explicitly with dedicated language constructs~\cite{CF22}, further reducing the amount of boilerplate code to be written in concrete development scenarios;
    \item the semantics are not implemented through a dedicated DSL, instead they can be written in \textsf{Java}, \textsf{Scala}, or other common general purpose programming languages for the JVM\@.
\end{enumerate}
Due to these three reasons, we believe Neverlang to be a suitable entry-level language workbench, as the only features that need to be learned concern the development of a grammar and the composition mechanisms, whereas developers do not need to learn a new language to implement the semantics.
For instance \textsf{Spoofax} uses \textsf{Stratego} to perform transformations over the AST through tree rewriting \textit{rules} and \textit{strategies}.
Therefore, we believe to be worth investigating how quick an unexperienced programmer can pick up a language workbench if the focus is put on the development paradigm rather than on the specific language constructs and their syntax.

\begin{figure}
    \centering
    \includegraphics[width=1\linewidth]{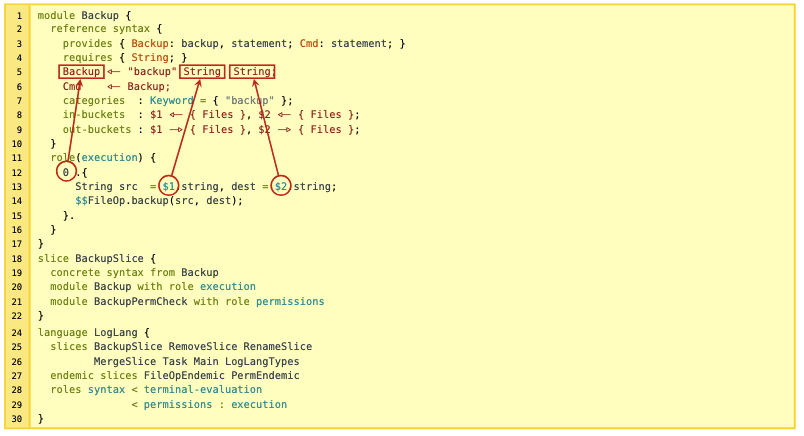}
    \caption{Exemplary language fragment in Neverlang.}
\label{fig:backup}
\end{figure} 

%\begin{Listing}[t]
   % \begingroup
   % \catcode`\!=\active%
   % \def!#1!{\tikz[na]\coordinate(#1);\label{#1}}
%   \setminted[neverlang]{escapeinside=@@}
%   \showneverlang*{Backup.nl}\vskip -10pt%
%   \caption{Exemplary language fragment in Neverlang.}\label{lst:backup}
   % \endgroup
%   \newcommand{\nt}[1]{%
%      \node[draw, fill=none, shape=circle, inner sep=0pt, minimum width=.375cm] (c\i) at (s\i) {};%
%      \coordinate (d\i0p) at ($(d\i0)+(-1pt,2pt)$);% chktex 1
%      \coordinate (d\i1p) at ($(d\i1)+(1pt,-3pt)$);% chktex 1
%      \node[draw, thick, shape=rectangle, name=d\i, inner sep=1pt, fit=(d\i0p) (d\i1p)] {};% chktex 1 % chktex 36
%   }
%   \begin{tikzpicture}[overlay, thick, BloodRed]
%      \coordinate (s0) at ($(s0)+(2pt,0pt)$) ;
%      \foreach \i in {0, 1, 2} {%
%         \nt{\i}
%         \draw[-stealth, rounded corners=7pt] (c\i) -- (d\i);
%      }
%   \end{tikzpicture}
%\end{Listing}

Neverlang~\cite{Cazzola12c, Cazzola13e, Cazzola15c} is a grammarware language workbench for creating programming languages (both general-purpose and domain-specific) in a modular way.
It supports manifold language extension and composition capabilities~\cite{Cazzola23b}. Figure~\ref{fig:backup} illustrates most of the crucial Neverlang elements. The basic Neverlang components are called slices (lines 18--22), %(lines~\ref{rb:slice}--\ref{re:slice}), 
which embody the concept of language features that can be compiled, tested, and distributed separately. This allows the same artefacts to be shared among different language implementations without requiring recompilation, in accordance with the constraints set by the language extension problem~\cite{Leduc20}. 
Slices are the result of the composition of multiple modules, which may include syntax definitions (lines 2--10) %(lines~\ref{rb:refsyntax}--\ref{re:refsyntax}) 
and/or roles. Roles define compilation phases by specifying semantic actions (lines 12--15) %(lines~\ref{rb:action}--\ref{re:action}) 
to be executed when certain syntax is recognised, as prescribed by the syntax-directed translation technique~\cite{Aho86}. The connection between the semantics and the respective syntax uses increasing positional numbers, as symbolised by the red arrows in Figure~\ref{fig:backup}. Eventually, all language fragments are composed together into a complete compiler through the language construct (lines 24--30).%(lines~\ref{rb:lang}--\ref{re:lang}).

%Neverlang natively supports LPL engineering thanks to AiDE~\cite{Cazzola20}.
\section{The Family of Experiments}\label{sec:studyDesign}

In the field of empirical software engineering, replications are an important factor in enhancing the validity of results~\cite{shull2008role, kitchenham2008role}.
The notion of replication has been extended to encompass the concept of \textit{family of experiments}~\cite{basili1999building}. A family comprises multiple similar experiments united in their scope to accumulate the necessary knowledge for deriving more meaningful conclusions with respect to a single experiment.

In this context, the experiments we conducted are best classified as quasi-experiments. Due to practical constraints, full randomisation of participants was not feasible; instead, subjects were selected opportunistically. This limitation is common in software engineering research, where participant access is often shaped by factors such as classroom availability or opportunities for professional collaboration~\cite{lenarduzzi2021towards}. Quasi-experimental designs are therefore frequently employed in empirical software engineering when full experimental control is not possible. Despite these limitations, such designs can still yield valuable insights---provided that threats to validity are carefully considered and mitigated~\cite{kampenes2009systematic}.

Before presenting the research questions, we first introduce the underlying theoretical framework used in our study---the tailored MEM---which provides the necessary foundation for understanding the key concepts explored in the research questions.

\begin{figure}
    \centering
    \includegraphics[width=0.5\linewidth]{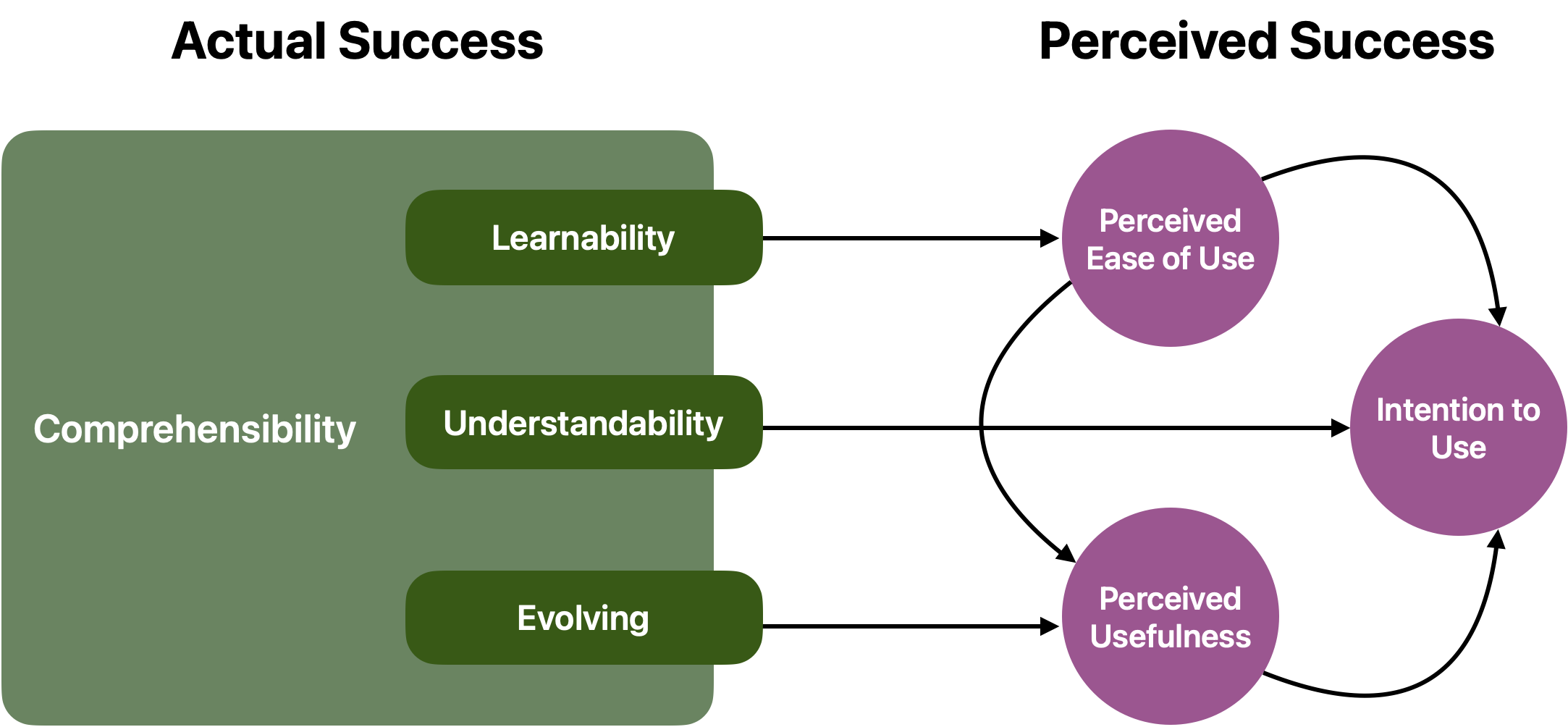}
    \caption{Tailored MEM. Actual success is expected to impact perceived success. Actual success is used as a proxy for comprehensibility, which is further decomposed into learnability, understandability, and evolving. Perceived success is decomposed into perceived ease of use, perceived usefulness, and intention to use.}
\label{fig:MEM}
\end{figure}

\subsection{Reference Theory: Tailored MEM}
To assess user-centred aspects such as the comprehensibility of the Neverlang meta-language and users' acceptance of the tool within the family of experiments, we employ the MEM~\cite{moody2001dealing}, a model used to evaluate information systems design methods, extending the Technology Acceptance Model (TAM)~\cite{davis1989perceived} by incorporating performance measures. Essentially, MEM integrates the concepts of \textit{perceived success} and \textit{actual success} to predict the real usage of a method.
We adapted the MEM to our context, introducing fine-grained measures of actual success. Figure~\ref{fig:MEM} illustrates the tailored MEM, and in the following we describe its different components.

\textbf{Actual Success. } Actual success entails evaluating the extent to which users effectively utilise the method, by means of \textit{performance-based} variables. In line with the methodology outlined in~\cite{abrahao2011evaluating}, we tailor MEM by measuring user performance in terms of their effectiveness in a test comprising a set of syntax and semantic understanding tasks. The test also involves the modification of a DSL excerpt expressed in the Neverlang meta-language. The measured effectiveness is used as a proxy to evaluate the comprehensibility of Neverlang.

We further adapt MEM by introducing a distinction between fine-grained and coarse-grained comprehensibility. Fine-grained comprehensibility delves into three distinct dimensions: \textit{learnability}, \textit{understandability}, and \textit{evolving}. Coarse-grained comprehensibility groups the different dimensions into a single measure---average across the dimensions.

\textit{Learnability} refers to the ability of learning the notation and meaning of the programs (i.e., the syntax of the language); \textit{understandability} refers to the ability of understanding the programs (e.g., identification of the correct meaning of the program, of language constructs, of new constructs, etc.); \textit{evolving} refers to the ability of modifying a program according to some needs.

Fine-grained dimensions serve as metrics for code usability and maintainability\footnote{Defined in ISO/IEC 25010:2011, \textit{Systems and software engineering — Systems and software Quality Requirements and Evaluation (SQuaRE) — System and software quality models}. Accessed online: 2024-08-02. Available at \url{https://www.iso.org/standard/78176.html}}, and strongly correlate with users' capacity to anticipate program behaviour~\cite{kosar2012program}.
We chose these dimensions not only because they have been adopted in several studies evaluating programming languages~\cite{kosar2012program, kosar2010comparing, kosar2009influence}, but also because they reflect the typical stages of learning a new language---grasping its syntax, understanding its semantics, and progressively modifying or extending code.

\textbf{Perceived Success. } Perceived success involves assessing users' perceptions of the method's quality.
It is measured through a combination of three \textit{perception-based} variables: \textit{perceived ease of use} (PEOU), denoting how user-friendly and accessible the method is considered; \textit{perceived usefulness} (PU), indicating the perceived suitability of the method for its intended practical use; and \textit{intention to use} (ITU), reflecting the user's inclination to utilise the method. These variables are evaluated through a specific MEM questionnaire~\cite{moody2001dealing}.

Overall, with the tailored MEM we aim to identify whether a relation exists between Neverlang comprehensibility (actual success), and language acceptance by its users (perceived success).

\subsection{Research Questions}
The goal of our family of experiments is to answer the following research questions (RQs):

\begin{description}
    \item[RQ1] \textit{To what extent is Neverlang comprehensible for novice users?} This RQ aims to understand the level of comprehension of Neverlang constructs and programs by users. The RQ is addressed by asking users to perform a skill test that assesses their comprehension of the language, and by measuring their performance in the test.
    \item[RQ2] \textit{To what extent is Neverlang accepted by novice users?} This RQ aims to understand how much users perceive the notation as easy to use and useful, and to what extent they intend to use the tool in the future. This is assessed by asking users to fill out a tailored MEM questionnaire.
    \item[RQ3] \textit{What is the relationship between Neverlang ease of use, usefulness, and intention to use it in the future?} This RQ aims at checking whether there is a relationship among the acceptance variables, and in particular, if ease of use and usefulness are related to intention to use.
    \item[RQ4] \textit{Is there a relationship between Neverlang comprehensibility and its acceptance by users?} This RQ aims to understand whether users who perform best in understanding Neverlang also tend to evaluate the tool as easier and more useful, and intend to use the tool in the future. To answer this question, the relationship between the variables associated with comprehensibility (measured through a test, cf.\ RQ1) and acceptance (measured through a questionnaire, cf.\ RQ2) is evaluated.
\end{description}

Since Neverlang is a novel tool that has not yet seen widespread adoption, our focus is on novice users---those with no prior experience with the tool. This allows us to evaluate its comprehensibility and acceptability from the perspective of first-time users, ensuring that the tool is accessible and usable for new adopters.

\subsection{Variables and Materials}
The variables based on the adapted MEM model (cf.\ Fig.~\ref{fig:MEM}) and the material used to collect them are defined below and summarised in Table~\ref{tab:variables}.
All the material used during the three experiments is publicly available online~\cite{broccia_2025_16094518}.

\subsubsection{Actual Success}
We distinguish between fine-grained comprehensibility, which considers the three different cognitive dimensions of comprehensibility separately (learnability, understandability, and evolving), and coarse-grained comprehensibility, which measures the average across the dimensions.

The evaluation of comprehensibility is based on the results obtained by the participants in a test.
The test contains questions that cover each of the three cognitive dimensions, following the template presented in~\cite{kosar2012program} (cf.\ Table~\ref{tab:test}).
The questions were defined in collaboration with two Neverlang experts, who considered it sufficiently effective to evaluate novices' comprehension of the language.

The evaluation of comprehensibility is based on the results obtained by the participants in a test.
The test includes questions covering each of the three cognitive dimensions, following the template proposed in~\cite{kosar2012program} (cf.\ Table~\ref{tab:test}).
The entire question development process was carried out in collaboration with two Neverlang experts. Starting from the template in~\cite{kosar2012program}---originally used to assess whether DSLs are more comprehensible than general-purpose languages---each question was carefully adapted to target key aspects of the Neverlang meta-language.
This adaptation involved several challenges, such as redefining what constitutes a ``program'' or a ``program result'' in the context of a meta-language---where the program describes how a language behaves, and its execution leads to the construction or modification of a language, rather than producing a direct runtime output.

\begin{figure}[t]
\caption{Example of a test question.}
\label{fig:q7}
\begin{tcolorbox}
\noindent
\textbf{7. Select from the options shown below, the one that indicates the correct meaning for the module shown in the image below.}

\vspace{0.5em}
\noindent
\begin{minipage}{0.55\textwidth}
    \includegraphics[width=\linewidth,keepaspectratio]{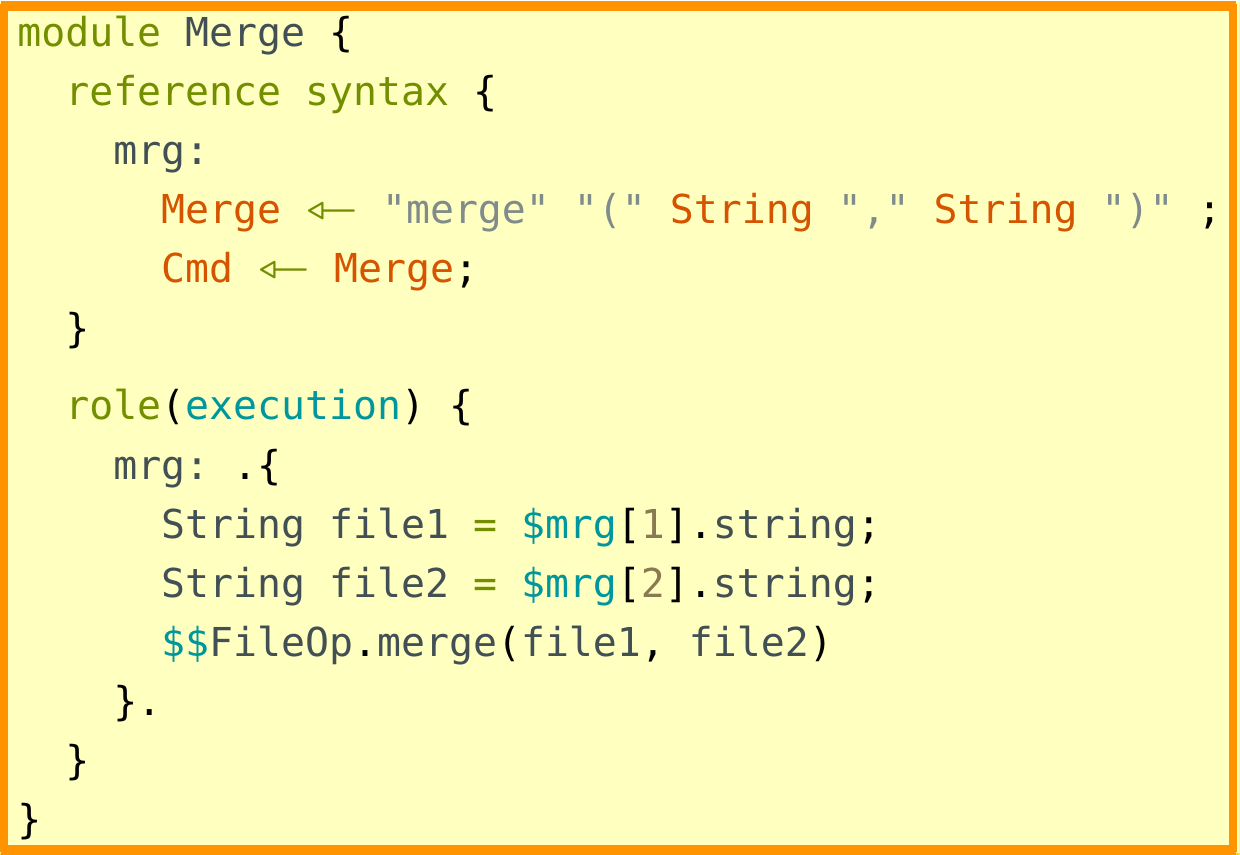}
\end{minipage}%
\hfill
\begin{minipage}{0.42\textwidth}
    \begin{itemize}[leftmargin=*,label=$\bigcirc$]
        \item\small{The \lq\lq Merge\rq\rq\ module adds a semantic role to "Task".}
        \item\small{The \lq\lq Merge\rq\rq\ module adds to the language the possibility of renaming a file with the merge command.}
        \item\small{The \lq\lq Merge\rq\rq\ module adds to the language the construct to create a new file from an old one.}
        \item\small{The \lq\lq Merge\rq\rq\ module adds to the language a command which merges two files into one.}
        \item\small{The \lq\lq Merge\rq\rq\ module adds an abstract syntax for the merge construct to the language.}
    \end{itemize}
\end{minipage}
\end{tcolorbox}
\end{figure}

The questions regarding \textit{learnability} and \textit{understandability} are multiple-choice questions and can contain a (set of) Neverlang statement(s), a description of a (set of) statement(s), or a simple request (e.g., select the set of Neverlang statements that satisfy a given characteristic). Each of these questions contains five exclusive choices in terms of sets of statements, statements' meanings, or given answers. Figure~\ref{fig:q7} shows question 7, as an example.

The questions regarding the \textit{evolving} dimension present a Neverlang program and a request for each of them. Users have to modify the program according to the request (e.g., remove from the Neverlang program all the constructs that manage a given characteristic).
For these questions, we evaluate answers as correct, partially correct, or incorrect based on assessments by two Neverlang experts. A correct answer receives a score of 1, a partially correct answer receives a score of 0.5, and an incorrect answer receives a score of 0.

Comprehension (both coarse- and fine-grained) is computed as the average of correct answers divided by the number of questions. The literature supports measuring comprehensibility by assessing the percentage of correct answers among different problem-solving tasks~\cite{abrahao2011evaluating, abrahao2012assessing, oliveira2020evaluating}.
We select a threshold of $0.6$, which corresponds to 60\% of answers being correctly answered; above such threshold, we consider comprehensibility as sufficient based on the academic grading standards in Italy~\cite{enwiki:1203203173}, which is the nation in which the experiments were conducted.  This choice ensures that the test aligns with local cultural and educational norms, where 6 out of 10 is widely recognised as the minimum acceptable performance. The questions were defined by Neverlang experts, who are also familiar with the Italian educational system. The experts agreed that correctly answering the 60\% of the questions would indicate a sufficient understanding of the tool's meta-language.

\begin{table*}[t]
\caption{Summary of variables.\\[-.25em]
{\scriptsize Variables labelled with $\circledast$ are measured only in the second replication of the experiment.}}
%\smallskip
%\begin{scriptsize}
\renewcommand{\arraystretch}{1.2}
\begin{tabular}{p{2.3cm} p{3.8cm}  p{2.8cm} p{5cm}}  \hline
%\hline
\textbf{Construct} & \textbf{Variables}  & \textbf{Material} & \textbf{Measure}\\
\hline
\multirow{4}{2.3cm}{Comprehensibility} & 1. Total Comprehensibility & \multirow{4}{2.8cm}{Test (cf.\ Table~\ref{tab:test})} & \multirow{4}{5cm}{Computed as the average proportion of correct test answers over the total number of questions}\\
                  & 2. Learnability  &      &              \\
                  & 3. Understandability  &        &         \\
                  & 4. Evolving  $\circledast$ &            &     \\ \hline
\multirow{3}{2cm}{Acceptance} & 1. Perceived Ease of Use (PEOU) & \multirow{3}{2.8cm}{Questionnaire (cf.\ Table~\ref{tab:perc-stat})} &  \multirow{3}{5cm}{Computed as the median of the questionnaire's statements points} \\
&   2. Perceived Usefulness (PU)  & & \\
 & 3. Intention to Use (ITU) &  &\\ \hline
\end{tabular}
\label{tab:variables}
%\end{scriptsize}
\end{table*}

\begin{table*}
    \caption{Questionnaire Template for the Comprehension Test.}
    \centering
\begin{tabular}{c c} \hline
   \multicolumn{2}{c}{\bfseries\em Learnability}      \\\hline
   Q1                                                   &        Select syntactically correct statements.                   \\
   Q2                                                   &        Select program statements with no sense (unreasonable).    \\
   Q3                                                   &        Select a valid program with the given result.              \\ \hline
   \multicolumn{2}{c}{\bfseries\em Understandability} \\\hline
   Q4                                                   &        Select the correct result for the given program.           \\
   Q5                                                   &        Identify language constructs.                              \\
   Q6                                                   &        Select a program with the same result.                     \\
   Q7                                                   &        Select the correct meaning for the new language construct. \\
   Q8                                                   &        Identify the correct meaning in the program with comments. \\ \hline
   \multicolumn{2}{c}{\bfseries\em Evolving}          \\\hline
   Q9                                                   &        Expand the program with new functionality.                 \\
   Q10                                                  &        Remove functionality from the program.                     \\
   Q11                                                  &        Change functionality of the program.                       \\\hline
\end{tabular}

\label{tab:test}
\end{table*}

\subsubsection{Perceived Success}
Users' acceptance is assessed by the three perception-based variables from the MEM: PEOU, PU, and ITU.
We gauge the three variables using a questionnaire consisting of 8, 14, and 8 statements, for each variable, respectively.
To mitigate systematic response bias, we employ a technique wherein the statements and their negated versions are included and shuffled (e.g., both ``Neverlang is simple and easy to understand'' and ``Neverlang is not simple and easy to understand'' are included). Respondents are required to assess each statement on a Likert scale ranging from 1 (\textit{strongly agree}) to 5 (\textit{strongly disagree}). Table~\ref{tab:perc-stat} presents the list of positive statements for PEOU, PU, and ITU.

Each variable is computed as the median of the points attributed to its statements (with points for negative statements considered as 6 minus the points provided by the respondent). Median is the appropriate descriptive statistics for ordinal data~\cite{harpe2015analyze}. We consider the value 3 (i.e., neither agree nor disagree) as the threshold above which user evaluations are categorised as positive.

\begin{table*}[t]
\renewcommand{\arraystretch}{1.2}
\caption{Perception-based statements (positive statements).}
%\smallskip
\begin{tabular}{c c c} \hline
%\setlength{\tabcolsep}{3pt}
%\resizebox{\columnwidth}{!}{\begin{tabular}{p{.7cm}| r  p{7.5cm}} \toprule
       &    & \textbf{Statements}  \\\hline

\textbf{PEOU}   & 1. & It was easy for me to understand what the code in Neverlang meant.                                         \\
       & 2. & Neverlang is simple and easy to understand.                                                                \\
       & 3. & Neverlang is easy to learn.                                                                                \\
       & 4. & Overall, Neverlang was easy to use.                                                                        \\\hline

\textbf{PU}     & 1. & Overall, I think Neverlang provides an effective means for defining a domain-specific language.            \\
       & 2. & I believe Neverlang has enough expressiveness to define a domain-specific language.                        \\
       & 3. & Overall, I find Neverlang to be useful.                                                                    \\
       & 4. & I think Neverlang is useful for defining a domain-specific language.                                       \\
       & 5. & Using Neverlang would improve my performance in defining a domain-specific language.                       \\
       & 6. & I believe Neverlang's code is organised, clear, concise, and unambiguous.                                  \\
       & 7. & I believe that using Neverlang would reduce the time needed to define a domain-specific language.          \\\hline

\textbf{ITU}    & 1. & If I were to work for a company in the future, I would use Neverlang to define a domain-specific language. \\
       & 2. & I intend to use Neverlang in the future if given the opportunity.                                          \\
       & 3. & I would recommend using Neverlang to programming language professionals.                                   \\
       & 4. & It would be easy for me to become proficient in using Neverlang.                                           \\
       \hline
\end{tabular}%
\label{tab:perc-stat}
\end{table*}

\subsection{Hypotheses}
To answer the RQs, we test a number of NULL hypotheses (cf.\ Table~\ref{table:tests}). The hypotheses for RQ1 check the coarse-grained (\textbf{H}$\bm{1_0}$) and fine-grained (\textbf{H}$\bm{2_0}$-\textbf{H}$\bm{4_0}$) comprehensibility of Neverlang. Those for RQ2 check its acceptance, and those for RQ3 check the relationships between acceptance variables.
The ones for RQ4 check a set of relationships between comprehensibility variables and acceptance variables, both coarse-grained (\textbf{H}$\bm{11_0}$-\textbf{H}$\bm{13_0}$), and fine-grained (\textbf{H}$\bm{14_0}$-\textbf{H}$\bm{22_0}$).
Not all the hypotheses are tested in all the experiments (we refer to Section~\ref{sect:individualExp} for the details on the individual experiments).
For the sake of conciseness, in Table \ref{table:tests} we report only the NULL hypotheses. The alternative hypotheses can be inferred by transposing them into positive form.

\begin{table}[t]
\renewcommand{\arraystretch}{1.2}
\caption{Hypotheses for each research question.\\ {\scriptsize* H1$_0$ focuses on coarse-grained comprehensibility \textcolor{black}{(i.e., total comprehensibility)}. \\ **H2$_0$, H3$_0$, and H4$_0$ aim at analysing fine-grained comprehensibility and focus on learnability, understandability and evolving, respectively.\\
Hypotheses labelled with $\circledast$ are tested only in the second replication of the experiment.}}
%\smallskip
\centering
\setlength{\tabcolsep}{4pt}
\begin{tabular}{|l|l|p{12cm}|}
\hline
%COMPRHENSIBILITY RQ1
\multirow{4}{*}{RQ1} & \textbf{H}$\bm{1_0}$ & Users are not effective in comprehending Neverlang constructs/programs *\\ \cline{2-3}
& \textbf{H}$\bm{2_0}$ & Users are not effective in learning Neverlang constructs/ programs **\\ \cline{2-3}
 & \textbf{H}$\bm{3_0}$ & Users are not effective in understanding Neverlang constructs/programs **\\ \cline{2-3}
 &     \textbf{H}$\bm{4_0}$~~~$\circledast$ & Users are not effective in modifying Neverlang constructs/ programs **\\  \hline

%ACCEPTANCE RQ2
\multirow{3}{*}{RQ2} & \textbf{H}$\bm{5_0}$ & Neverlang is perceived as difficult to use \\ \cline{2-3}
 &     \textbf{H}$\bm{6_0}$& Neverlang is perceived as not useful \\ \cline{2-3}
 &     \textbf{H}$\bm{7_0}$ & There is no intention to use Neverlang in the future \\ \hline

%ACCEPTANCE RELATIONSHIP RQ3

\multirow{3}{*}{RQ3} &    \textbf{H}$\bm{8_0}$ & There is no relationship between perceived ease of use and perceived usefulness  \\ \cline{2-3}
 &    \textbf{H}$\bm{9_0}$ & There is no relationship between perceived usefulness and intention to use \\ \cline{2-3}
 &    \textbf{H}$\bm{10_0}$ & There is no relationship between perceived ease of use and intention to use   \\ \hline

% COMPREHENSIBILITY AND ACCEPTANCE RQ4
\multirow{11}{*}{RQ4} &    \textbf{H}$\bm{11_0}$ & There is no relationship between  \textcolor{black}{total} comprehensibility  and perceived ease of use      \\ \cline{2-3}
 &     \textbf{H}$\bm{12_0}$ & There is no relationship between \textcolor{black}{total} comprehensibility and perceived usefulness \\ \cline{2-3}
 &     \textbf{H}$\bm{13_0}$ & There is no relationship between  \textcolor{black}{total} comprehensibility and intention to use \\ \cline{2-3}
 &    \textbf{H}$\bm{14_0}$ & There is no relationship between learnability and perceived ease of use      \\ \cline{2-3}
  &     \textbf{H}$\bm{15_0}$ & There is no relationship between learnability and perceived usefulness \\ \cline{2-3}
  &     \textbf{H}$\bm{16_0}$ & There is no relationship between learnability and intention to use \\ \cline{2-3}
 &    \textbf{H}$\bm{17_0}$ & There is no relationship between understandability and perceived ease of use      \\ \cline{2-3}
  &     \textbf{H}$\bm{18_0}$ & There is no relationship between understandability and perceived usefulness \\ \cline{2-3}
  &     \textbf{H}$\bm{19_0}$ & There is no relationship between understandability and intention to use \\ \cline{2-3}
 &     \textbf{H}$\bm{20_0}$~$\circledast$ & There is no relationship between evolving and perceived ease of use      \\ \cline{2-3}
 &      \textbf{H}$\bm{21_0}$~$\circledast$ & There is no relationship between evolving and perceived usefulness \\  \cline{2-3}
 &      \textbf{H}$\bm{22_0}$~$\circledast$ & There is no relationship between evolving and intention to use \\
 \hline

\end{tabular}

\label{table:tests}
\end{table}

\subsection{Experiment Phases}
Both the original experiment and the two replications are composed of the following four phases:

\begin{itemize}

    \item \textbf{Training phase:} Users are required to undergo training on Neverlang, equipping them with all requisite information to complete the test. In UniMI 2022 and UniMi~2 2023, the training was delivered through a video (cf.\ Section~\ref{sec:originalexp}), while in BISS 2023, it consisted of a seminar (cf.\ Section~\ref{sec:firstreplication}). 
    \item \textbf{Test phase:} Users are asked to answer the questions about Neverlang on an online survey platform. The questions are divided according to the comprehensibility dimensions (cf.\ Table~\ref{tab:test} for the questionnaire template). For each dimension, the questions were presented with increasing levels of difficulty. This structure follows the approach used by Kosar et al.~\cite{kosar2012program} and is in line with gamification principles~\cite{dicheva2015gamification}. By gradually increasing the difficulty of the questions, we ensure that participants first establish a basic understanding before tackling more complex aspects, potentially leading to more reliable responses. Additionally, this structured approach helps mitigate the risk of cognitive overload, allowing users to adapt and build confidence as they progress.
    \item \textbf{Post-study phase:} Users are asked to fill out an online questionnaire containing a number of questions about their prior knowledge (on programming and language workbenches), personal details (age, gender, occupation), and feedback on the test and on the training material\footnote{The qualitative feedback was rather limited, and was not systematically analysed. However, we present some excerpts in the discussion, when these are relevant to support certain arguments.}. Prior knowledge was assessed by asking participants to self-evaluate on a scale from 1 (very low) to 5 (very high), while feedback was collected through open-ended questions regarding their opinions on Neverlang and their overall experience.
    The questionnaire also includes the 5-point scale questions on PEOU, PU, and ITU~\cite{moody2001dealing}.
\end{itemize}

The core part of the experiment, i.e., the test phase, was estimated to last approximately 1~hour. This duration was chosen to ensure that users could engage with the tasks attentively while maintaining a balance between thorough evaluation and cognitive strain. To prevent stress, the participants were not given a time limit, and they were recommended to interrupt the study in case they felt uncomfortable.
It should be noted that the experimental mortality is limited (21\%
in total), which confirms that the participants were comfortable with the time constraints and the overall task. Concerning privacy aspects, the tests were anonymised, and the experimenter never met the participants. This also mitigates a possible Hawthorne effect that may have emerged during the post-study phase,  concerned with acceptance variables.

\section{Individual Experiments}\label{sect:individualExp} 
In this section, we describe the characteristics of each of the three experiments that constitute our family of
experiments. 
Participants across all three experiments were recruited on a voluntary basis through in-person direct requests made during university lectures and doctoral school sessions, with no compensation provided for their involvement. They were informed that they could withdraw from the study at any time without consequence, and that the collected data would be used for scientific publication in an anonymous and aggregated form.

\subsection{Original Experiment: UniMI 1 2022}
\label{sec:originalexp}
\begin{figure}
\centering
\resizebox{0.8\linewidth}{!}{
\begin{tabular}{>{\centering\arraybackslash}p{5cm} >{\centering\arraybackslash}p{5cm} >{\centering\arraybackslash}p{5cm}}
    \textbf{\textsf{\textcolor{black!75}{Programming}}} & 
    \textbf{\textsf{\textcolor{black!75}{Java/Object-oriented}}} &
    \textbf{\textsf{\textcolor{black!75}{Experience with}}} \\
    \textbf{\textsf{\textcolor{black!75}{skills}}}
    &
    \textbf{\textsf{\textcolor{black!75}{programming skills}}}
    &
    \textbf{\textsf{\textcolor{black!75}{language workbenches}}}
    \\ & & \\
    \includegraphics[width=1\linewidth]{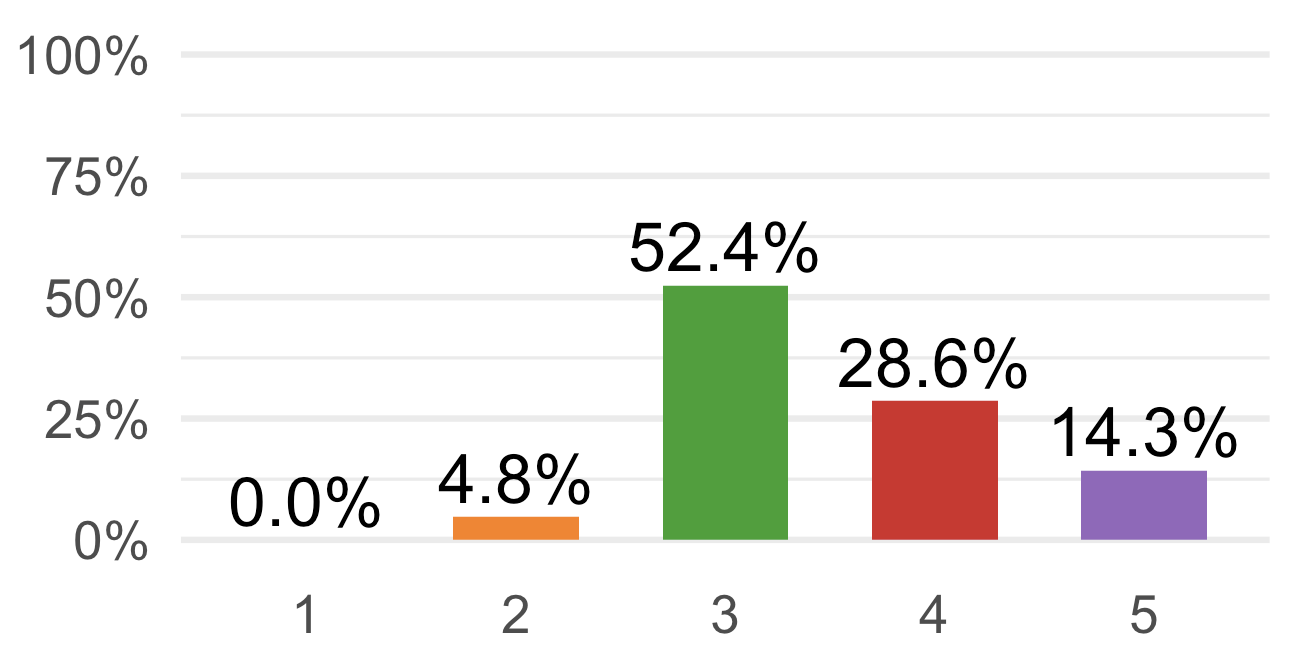}
     & 
     \includegraphics[width=1\linewidth]{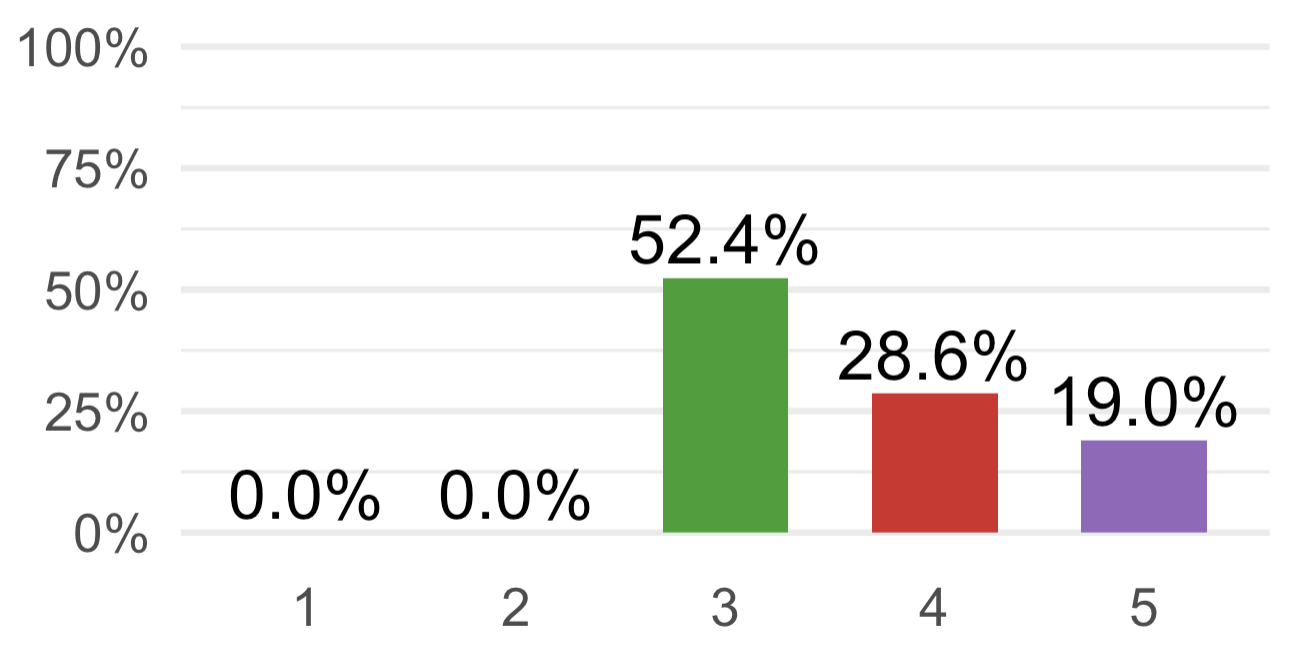}
     & 
     \includegraphics[width=1\linewidth]{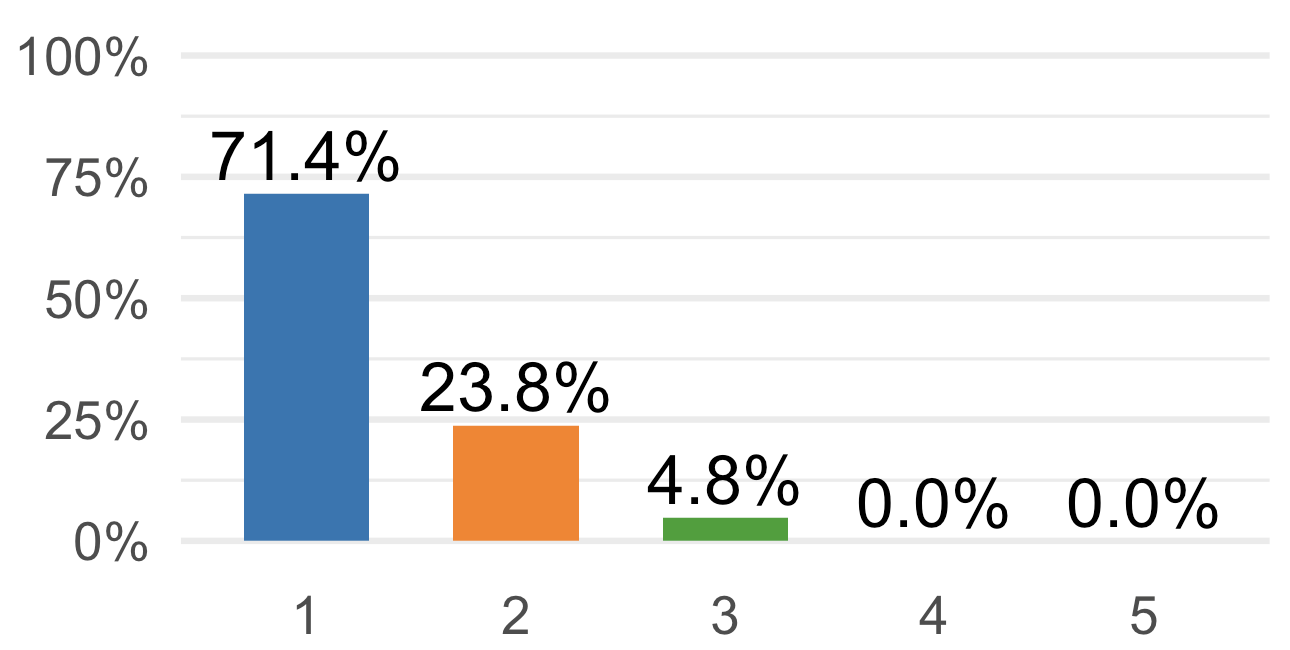}
     \\ \\ & & \\
     \multicolumn{3}{c}{\fontsize{11}{24}\selectfont (a) Original Experiment} \\ \\ & & \\
     \includegraphics[width=1\linewidth]{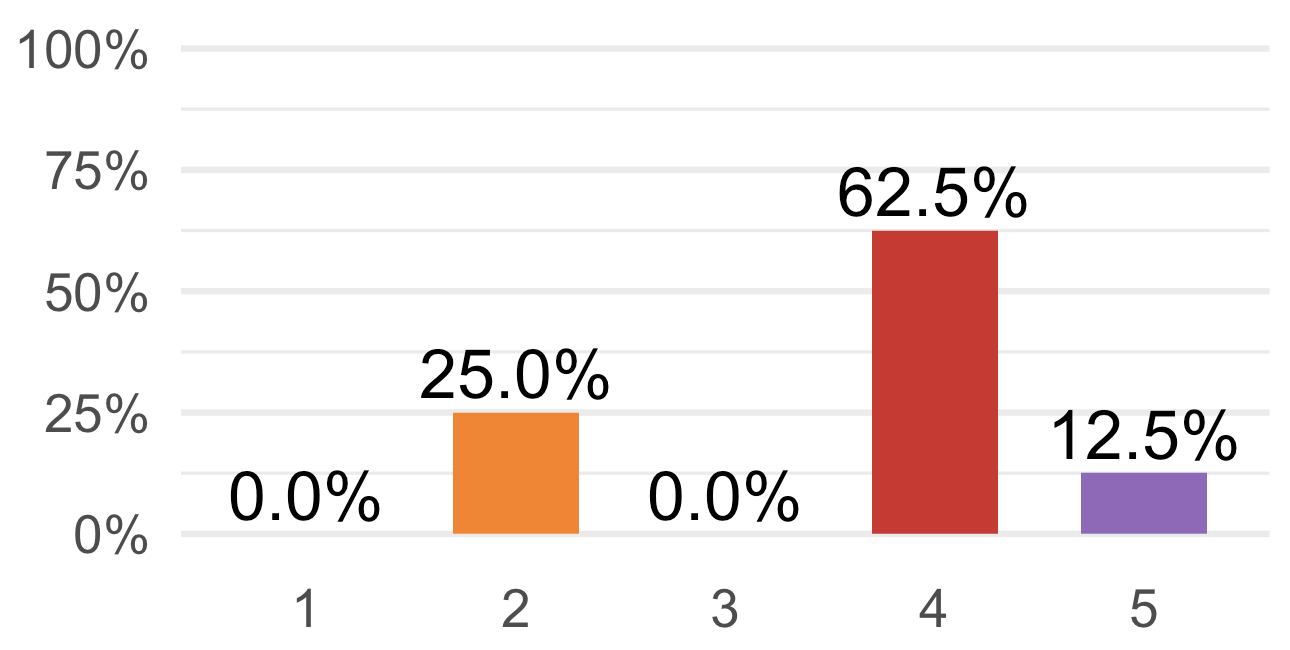}
     & 
     \includegraphics[width=1\linewidth]{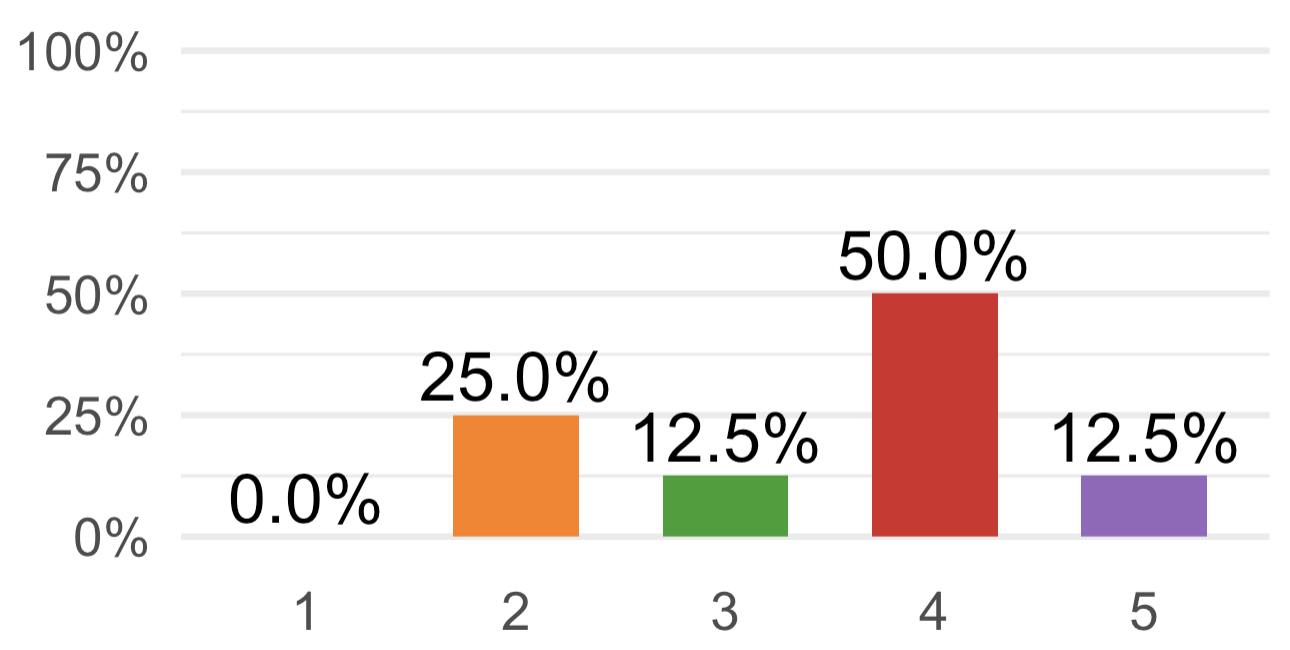}
     & 
     \includegraphics[width=1\linewidth]{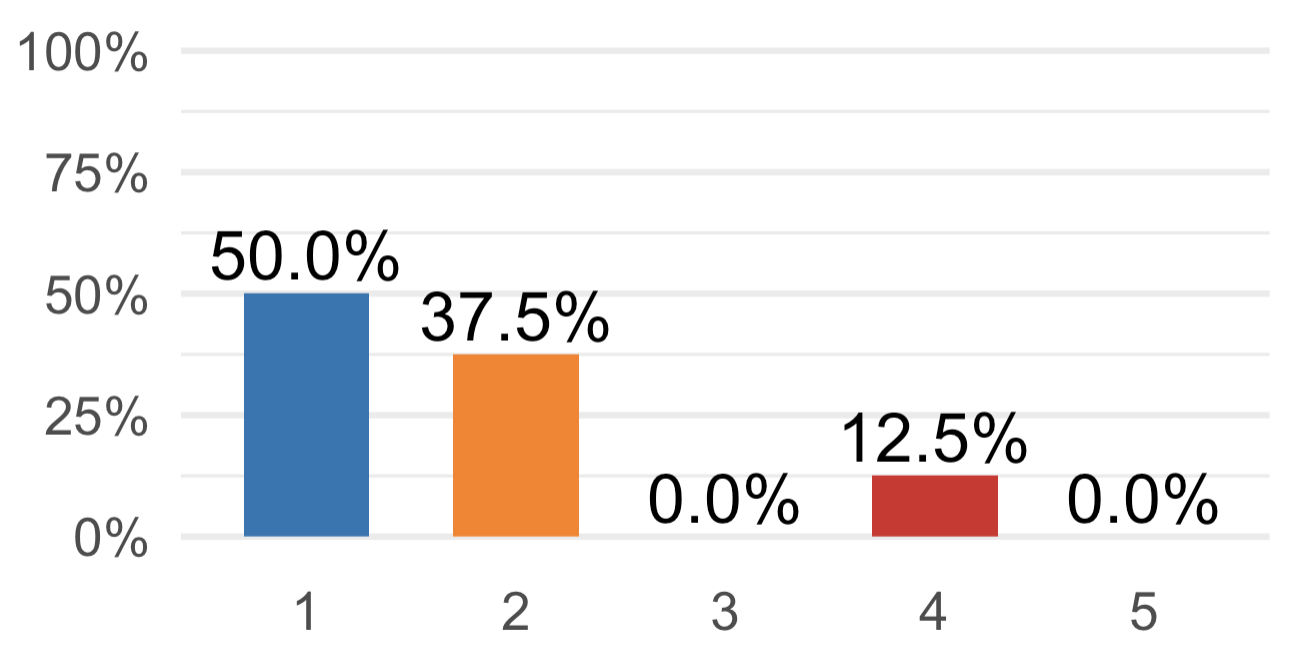}
     \\ & & \\
     \multicolumn{3}{c}{\fontsize{11}{24}\selectfont (b) First Replication} \\ \\ & & \\
     \includegraphics[width=1\linewidth]{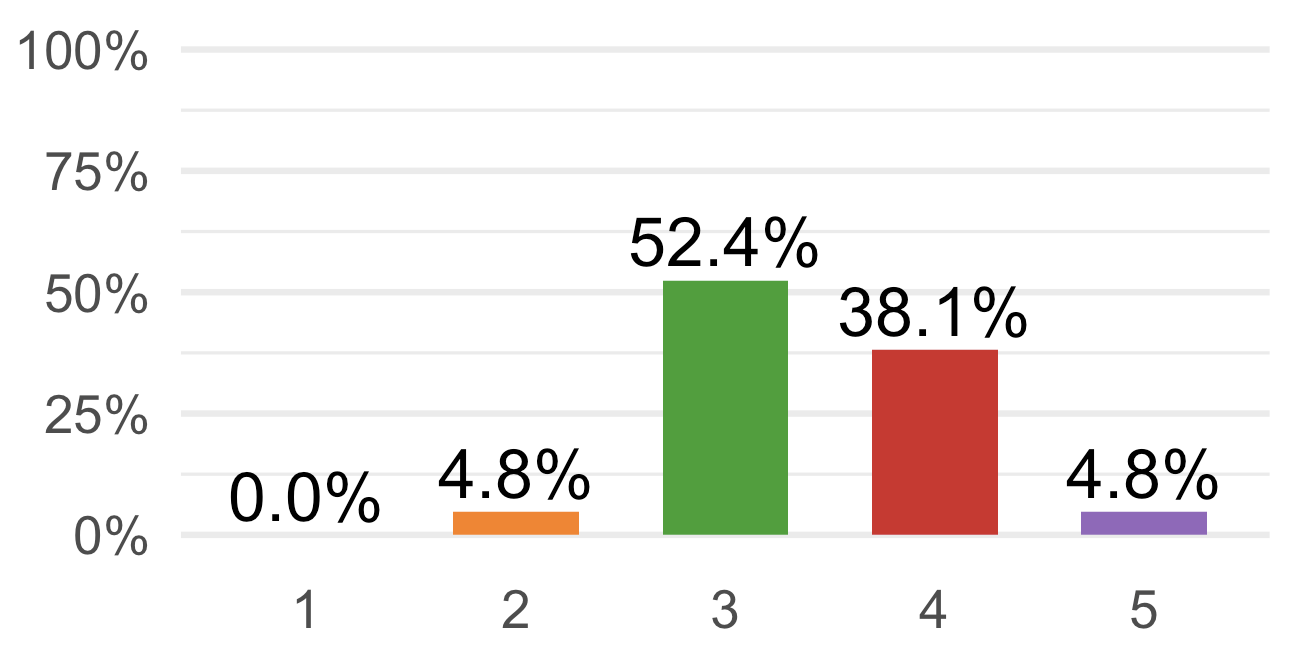}
     & 
     \includegraphics[width=1\linewidth]{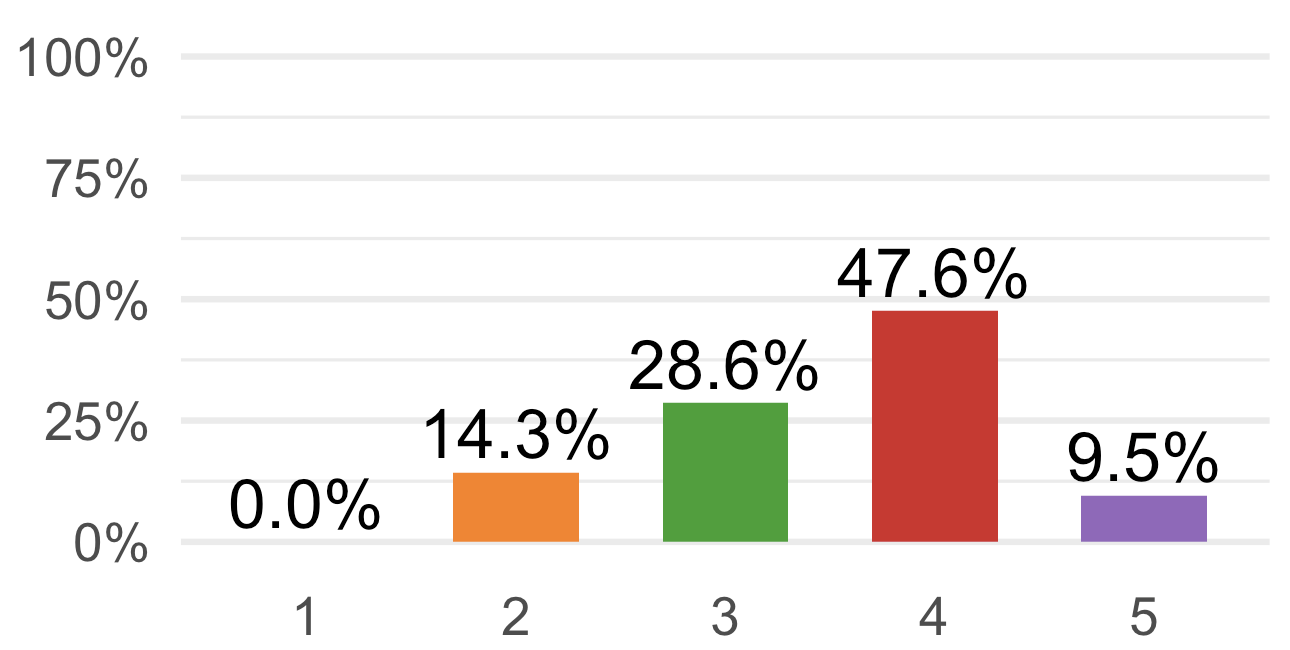}
     & 
     \includegraphics[width=1\linewidth]{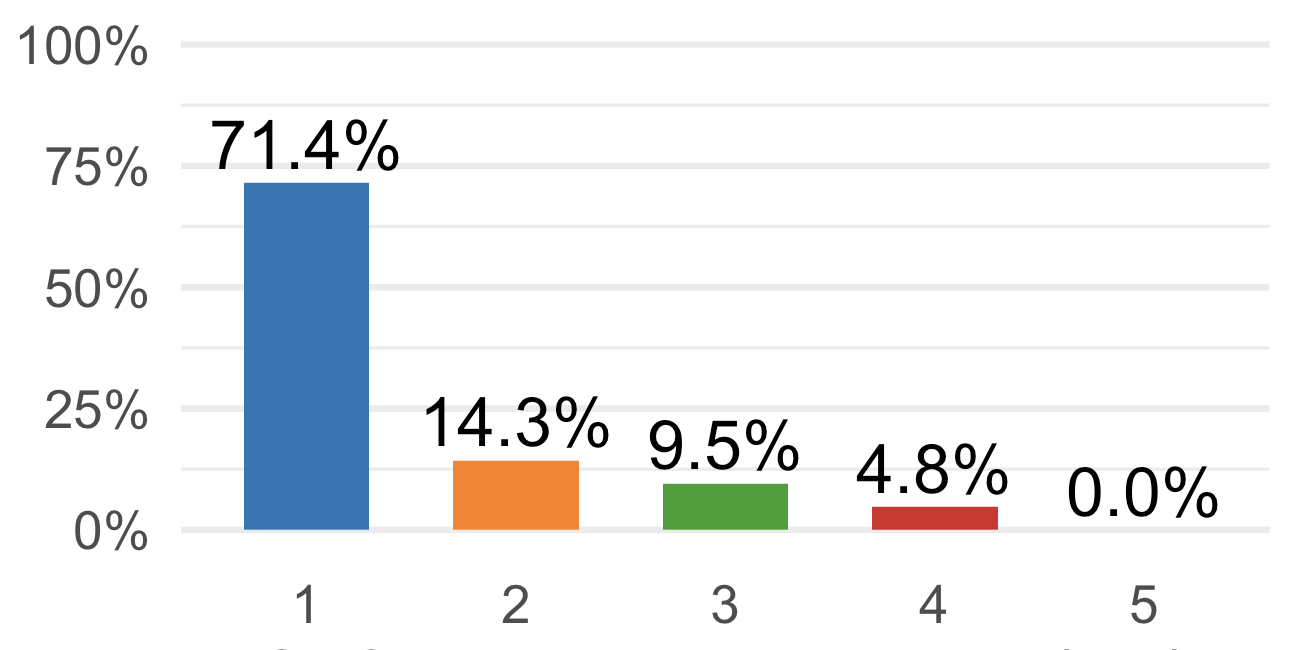}
     \\ & & \\
     \multicolumn{3}{c}{\fontsize{11}{24}\selectfont (c) Second Replication} \\
\end{tabular}
}
\caption{Self-evaluated participants' prior knowledge ranging from 1 (very low) to 5 (very high).} 
\label{fig:post1}
\end{figure}

\textbf{Participants. }
We recruited 21 subjects (9.5\% female, 90.5\% male; aged between 21 and 29 y.o.) from the University of Milan. The subjects are bachelor students~(9), master students~(10), graduated subjects~(1), and PhD students~(1) in Computer Science. Their skills in programming, in Java/object-oriented programming, and their experience with the language workbenches  are self-evaluated on a scale from~1 to~5. Figure~\ref{fig:post1} 
%and Table~\ref{tab:priorKnow} (cf. column Or.) 
reports the results. 

\medskip
\noindent\textbf{Phases and Material. }
The experiment was entirely conducted online during 2022, with all phases conducted in Italian, the native language of the participants.

The training phase was conducted through a video that presents Neverlang and gives users all the necessary information to perform the test phase. The video lasts less than 15 minutes and is available online\footnote{\url{https://youtu.be/q3E9MxmtOXE?si=M5ToYNawjCxYSdtb}}.  Participants were asked to watch the video before the test, preferably once. However, they can use this support during the test in case of need, provided they report the number of times they did so.

The comprehensibility test used was composed only of learnability and understandability questions. 

\medskip
\noindent\textbf{Hypotheses. } Since the comprehensibility test included only learnability and understandability questions, H4$_0$, H17$_0$, H18$_0$, H19$_0$, and H23$_0$ were not tested in this experiment.

\subsection{First Replication: BISS 2023}
\label{sec:firstreplication}
\textbf{Participants. }
We recruited 8 subjects (\textcolor{black}{12.5\% female, 87.5\% male}; aged between 24 and 37 y.o.) from BISS 2023, an annual doctoral school offering graduate-level courses for national and international PhD students. The participants were PhD students in Computer Science from the University of Pisa, University of Bologna, and the Italian National Research Council. 
Their self-evaluated prior knowledge on programming and language workbenches is reported 
%in Table~\ref{tab:priorKnow} (cf. column I Rep.) and 
in Fig.~\ref{fig:post1}.

\medskip

\noindent\textbf{Phases and Material. }
The experiment was conducted online, with the exception of the training phase, which was conducted as a 3-day seminar (12 hours) during the doctoral school, evenly split between theoretical lessons and hands-on sessions.
%\textcolor{black}{@UNIMI: aggiungere di quante ore.}
All the phases were conducted in English, even if it was not the native language of the participants.

In this case too, the comprehensibility test used was composed only of learnability and understandability questions. 

\medskip
\noindent\textbf{Hypotheses. } Since the comprehensibility test included only learnability and understandability questions, H4$_0$, H17$_0$, H18$_0$, H19$_0$, and H23$_0$ were not tested in this experiment.

\subsection{Second Replication: UniMI 2 2023}
\textbf{Participants. }
We recruited 21 subjects (\textcolor{black}{4.8\% female, 81\% male, 14.2\% preferred not to answer}; aged between 21 and 41 y.o.) from the University of Milan. 
The subjects are bachelor students (12) and master students (9) in Computer Science. 
Their prior knowledge is reported 
%in Table~\ref{tab:priorKnow} (cf. column II Rep.) and 
in Fig.~\ref{fig:post1}. 

\medskip
\noindent\textbf{Phases and Material. }
The experiment was conducted entirely online, with all phases carried out in Italian, the native language of the participants, except for two individuals.  %\textcolor{blue}{tutti madre lingua italiani tranne 2 una tedesca e l'altro nord africano ma non saprei bene dire di dove.}

The training phase was conducted through a video, as in the original experiment. 

The comprehensibility test used was composed of learnability, understandability, and evolving questions. 

\medskip
\noindent\textbf{Hypotheses. } All hypotheses were tested in this experiment.

\subsection{Differences between Experiments}
\label{sec:differences}

The original experiment and second replication are highly comparable; the first replication differs in several aspects. 

Regarding sample size, the first replication had a smaller group of 8 participants, compared to 21 participants in both the original experiment and the second replication. 

In terms of seniority level, the first replication involved only PhD students, while the original experiment included just one PhD student, and the second replication had none. This difference is visible in Fig.~\ref{fig:post1}, which shows that participants in the first replication had better programming skills and slightly more experience with language workbenches.

Additionally, the training phase varied between the experiments. In the original experiment and the second replication, training was conducted using a 15-minute video. In contrast, the first replication involved a more extensive 12-hour seminar for training.

The differences in demographics are due to our opportunistic sampling. The substantial difference in terms of training is due to the specific context of the first replication, i.e., a doctoral course. The potential impact of these aspects will be discussed when the results of the experiments are not fully aligned (cf.\ Section~\ref{sec:results}). However, we anticipate that most of the results are consistent across the experiments, which increases the generalisability of our findings to contextual  variations.

%======================= DATA ANALYSIS AND RESULTS =================

\section{Data Analysis and Results}
\label{sec:results}

In this section, we present the analysis procedure used to answer the RQs and the results for all experiments, as well as the meta-analysis conducted on the family of experiments. 
Table~\ref{table:resultsRQ} shows a summary of the responses to all RQs.

\subsection{Data Analysis Procedure}\label{sec:dataAnalysisProcedure}

\textbf{Statistical Tests.} To answer RQ1, we checked the dataset for normality using the Kolmogorov-Smirnov test, which resulted in p-values below the 0.05 significance level, indicating a failure of the normality test for all dependent variables~\cite{wohlin2012experimentation}. Due to this, we employed a non-parametric test, the Wilcoxon signed-rank test, to check whether comprehensibility (both coarse-grained and fine-grained) is significantly above the target value of $0.6$. Comprehensibility greater than $0.6$ is considered to indicate sufficient comprehension according to the Italian academic grade system. Additionally, we measured effect size using Cohen's D test~\cite{cohen1992statistical}, allowing us to interpret the magnitude of the effect in context. Cohen's D values are typically interpreted as small (0.2), medium (0.5), and large (0.8), where larger effect sizes imply more impactful effects~\cite{cohen1992statistical}. 

To answer RQ2, we applied a Wilcoxon signed rank test to check whether PEOU, PU, and ITU are significantly above the value of the Likert scale representing neutral perception (i.e., 3). As with RQ1, we apply a non-parametric test due to the failure of the normality test for all variables. To compute the effect size, we use Rank-Biserial correlation, which is suitable for non-parametric tests
with ordinal data~\cite{glass1966note}. The rank-biserial correlation ranges from $-1$ to $+1$, the closer the value is to $\pm$1, the stronger the effect.

To address RQ3 and RQ4 we employ Spearman's Rank Correlation test, a non-parametric test appropriate for non-normally distributed ordinal (acceptance) and proportional (comprehensibility) variables, given the failure of the normality assumption across all variables. 

For RQ3, we examine the relationships between PEOU and PU, PEOU and ITU, and PU and ITU. 
For RQ4, we investigate the relationship between comprehensibility and both PEOU and PU. This analysis is also extended to learnability, understandability, and evolving (in the second replication).
%For RQ5, we analyse the relationship between WM capacity and comprehensibility. In this case too, we test the relationship between WM capacity and both coarse-grained comprehensibility and fine-grained comprehensibility (i.e., learnability, understandability, and evolving).

In addition to the p-value, which determines the significance of the relationship, Spearman's Rank Correlation test provides the $\rho$ coefficient, which assesses whether the relationship between two variables is monotonic (i.e., as one variable changes, the other variable consistently increases or decreases). The coefficient effectively measures both the strength and direction of the relationship between the variables, ranging from -1 (indicating a perfect negative relationship) to +1 (indicating a perfect positive relationship).

\textbf{Correction for Multiple Comparisons.} Performing multiple statistical tests increases the likelihood of Type~I errors. A common method to control these errors is to apply a Bonferroni correction, which adjusts the significance level by dividing it by the number of tests. However, this method is often overly conservative, which can increase the risk of Type~II errors \cite{perneger1998s}. In our study, we used the Holm-Bonferroni correction, a preferred alternative that is both powerful and less conservative. This approach controls the family-wise error rate while reducing the risk of Type~II errors, providing a more balanced method for multiple comparisons \cite{barnett2022multiple}. We applied the Holm-Bonferroni correction separately to the set of hypotheses associated with each RQ.

The Holm-Bonferroni procedure begins by sorting the original p-values in ascending order (from smallest to largest). For each p-value $p_i$, an adjusted significance threshold is computed as: $\alpha_i = \frac{\alpha}{n - i + 1}$, where $\alpha$ is the overall significance level (i.e., 0.05), $n$ is the total number of tests (i.e., number of hypotheses), and $i$ is the rank of the p-value (1 for the smallest, 2 for the second smallest, etc.). 
To facilitate interpretation, p-values can be adjusted so they can be directly compared to the original significance level. The adjusted p-values are computed as: $Adjusted p_i = min(1, p_i \times ( n - i + 1))$, where $p_i$ is the i-th smallest p-value. These adjusted p-values provide a consistent way to report results, particularly when dealing with multiple comparisons, making it easier to determine statistical significance.

\textbf{Family Meta-analysis.} 
Finally, to aggregate and interpret the results from our family of experiments, we employed meta-analysis, a statistical technique widely used across disciplines such as medicine, psychology, epidemiology, and public health~\cite{wang2023conducting}. Among the various approaches, aggregated data (AD) meta-analysis is the most common in systematic literature reviews~\cite{kitchenham2004procedures}, offering both quantitative rigour and the ability to manage heterogeneity across studies~\cite{santos2018analyzing}.

In AD meta-analysis, the primary focus is on effect-size metrics (e.g., standardised mean difference and odds ratio) that quantify the relationship between experimental groups. Depending on the research question, different types of effect sizes can be used. In our study, we employed two types of meta-analyses: \textit{meta-analysis of proportions} (for RQ1 and RQ2), and \textit{meta-analysis of correlation coefficients} (for RQ3 and RQ4).
In meta-analysis of proportions, the effect-size metric represents the estimated overall proportion of individuals who meet a predefined criterion. In contrast, meta-analysis of correlation coefficients 
aggregates the strength and direction of relationships between pairs of variables across studies. 

For both types of meta-analyses, we adopted a random-effects model, which accounts for potential variability between studies. This choice is particularly appropriate in our context, given differences in participant populations, experimental procedures, and training formats across replications.

For RQ1 and RQ2, we conducted a meta-analysis of proportions to estimate the proportion of users exceeding the target value of 0.6 for comprehensibility, and 3 for acceptance measures (PEOU, PU, ITU). The results include the estimated overall proportion, its confidence interval, and an assessment of statistical significance. We also examined heterogeneity metrics (e.g., $I^2$) and visualised the results using forest plots, which summarise the proportion estimates and confidence intervals from each study along with the combined effect.

For RQ3 and RQ4, we conducted a meta-analysis of correlation coefficients using Fisher’s Z-transformation~\cite{van2023meta}. Correlation values (Spearman’s $\rho$) from individual studies were first transformed into Z-scores, then aggregated using a random-effects model to account for sampling variability. The combined effect was back-transformed to the correlation scale, and confidence intervals and p-values were computed to assess the strength and statistical significance of the associations between PEOU, PU, ITU, and the comprehensibility dimensions.

It is important to note that the interpretation of confidence intervals differs between the two types of meta-analyses. In meta-analysis of proportions, a neutral threshold of 0.5 is used: if the confidence interval includes 0.5, we cannot claim that a majority of users exceeded the target value, and the result is not statistically significant. In contrast, in meta-analysis of correlation coefficients, zero is the reference point: if the confidence interval includes 0.0, the association might be absent, and the result is not considered statistically significant.

\begin{table*}[t]
\renewcommand{\arraystretch}{1.2}
\caption{Summary of the responses to each Research Question.}
%\smallskip
\scriptsize
\begin{tabular}{|c|p{6cm}|p{7cm}|}
\hline
RQ1 &\textit{To what extent is Neverlang comprehensible for novice users?} &   \textcolor{black}{Users exhibit sufficient Neverlang comprehensibility, especially concerning its syntax.} \\\hline

RQ2 & \textit{To what extent is Neverlang accepted by users?} & 
\textcolor{black}{Users demonstrate a favourable perception of the tool usefulness and express an intention to use it in the future.}\\ \hline

RQ3 & \textit{What is the relationship between Neverlang ease of use/usefulness and intention to use it in the future?} &  \textcolor{black}{Variation in users' perceived ease of use and perceived usefulness  are associated with their intention to use Neverlang.}\\ \hline
RQ4  & \textit{Is there a relationship between Neverlang comprehensibility and its acceptance by users?}  & There is no discernible relationship between the comprehensibility of the Neverlang and users' acceptance of the tool.\\  \hline
%RQ5 & \textit{Is Neverlang more comprehensible for users with a higher working memory capacity?}  & There is no discernible relationship between users' working memory capacity and their comprehensibility of Neverlang.\\  \hline

\end{tabular}

\label{table:resultsRQ}
\end{table*}

\subsection{Results: Individual Experiments}

\textbf{Descriptive Statistics.} Table~\ref{tab:descrStats} shows descriptive statistics for all the variables gathered with the test for the original experiment and its two replications.

\begin{table*}[t]
    \renewcommand{\arraystretch}{1.2}
    \caption{Descriptive statistics.\\[-.25em] 
{\scriptsize Columns denoted with \textbf{Or.} report the results for the original experiment (UniMI 1 2022); columns denoted with \textbf{I r.} report the results for the first replication (BISS 2023); columns denoted with \textbf{II r.} report the results for the second replication (UniMi 2 2023).}}
    \label{tab:descrStats}
    %\smallskip
    \centering
    \scriptsize
    \vspace*{-0.05cm}
    \begin{tabular}{|c |c|c|c| c|c|c| c|c|c |c|c|c |c|c| c|}
    \hline 
      {\bf{\em Variables}} & \multicolumn{3}{c|}{\bf{\em Median}}  & \multicolumn{3}{c|}{\bf{\em Mean}} & \multicolumn{3}{c|}{\bf{\em Std.\,dev.}} & \multicolumn{3}{c|}{\bf{\em Min.}} & \multicolumn{3}{c|}{\bf{\em Max.}} \\ 
      \hline
      & \textbf{Or.} & \textbf{I r.} & \textbf{II r.} & \textbf{Or.} & \textbf{I r.} & \textbf{II r.}& \textbf{Or.} & \textbf{I r.} & \textbf{II r.}& \textbf{Or.} & \textbf{I r.} & \textbf{II r.}& \textbf{Or.} & \textbf{I r.} & \textbf{II r.}\\ \hline
    \textcolor{black}{Total} Comprehensibility & 0.8  & 0.817 & 0.756 & 0.724 & 0.779 & 0.686 & 0.249 & 0.195 & 0.185 & 0.1 & 0.467 & 0.289 & 1 & 1 & 0.944\\ \hline
    Learnability & 1  & 1 & 1 & 0.762 & 0.833 & 0.778 & 0.336 & 0.252 & 0.285 & 0 & 0.333 & 0 & 1 & 1 & 1 \\ \hline
    Understandability & 0.8  & 0.7 & 0.8 & 0.686 & 0.725 & 0.676 & 0.241 & 0.282 & 0.264 & 0.2 & 0.2 & 0 & 1 & 1 & 1 \\ \hline
    Evolving  & -  & - & 0.667  & -  & - & 0.603 &-  &-  & 0.291 &-  &-  & 0 &-  &-  & 1 
    \\   
      \hline
    PEOU & 3  & 3.25 & 3 & 3 & 3.125 & 3.167 & 0.816 & 0.641 & 0.764 & 2 & 2 & 1.5 & 4 & 4 & 4\\ \hline
    PU &  4 & 4 & 4 & 4.119 & 4.167 & 3.976 & 0.350 & 0.408 & 0.512 & 3.5 & 4 & 3 & 5 & 5 & 5\\ \hline
    ITU &  4 & 4 & 3.5 & 3.643 & 3.438 & 3.405 & 0.573 & 0.821 & 0.490 & 2.5 & 2 & 2.5 & 5 & 4 & 4 \\ \hline
    %WM Capacity & 0.819   & 0.823 & 0.816  & 0.807 & 0.814 & 0.818 & 0.109 & 0.119 & 0.084 & 0.584 & 0.579 & 0.621 & 0.992 & 1 & 0.926 \\ \hline
   \end{tabular}
\end{table*}

The results indicate a generally good level of comprehensibility with an average  \textcolor{black}{total} comprehensibility above $\sim$0.69 for all the experiments, meaning $\sim$$69\%$~of the questions of the test are correctly answered.
Regarding the different cognitive dimensions composing comprehensibility, the results show that learnability is the measure that provides the highest contribution in all the experiments (mean score 0.762, 0.833, and 0.778, respectively), followed by understandability (mean score 0.686, 0.725, and 0.676, respectively). The evolving measure was measured only in the second replication and shows a sufficient score. This suggests that while syntax is generally well understood by users, they tend to have more difficulty in understanding the meaning of the programs and modifying them (namely, in using the tool in practice).

The results show that among the acceptance variables (PEOU, PU, and ITU), PU consistently receives the highest ratings across all three experiments. In each case, the median score is 4, which is above the neutral midpoint (i.e., 3, representing "neither agree nor disagree"), suggesting a generally positive perception of the tool's usefulness.
Conversely, the least appreciated aspect is PEOU, with median scores of 3, 3.25, and 3, respectively, across the experiments. These values are close to or equal to the neutral point, indicating that participants were uncertain or ambivalent about the ease of use of the tool.
As for ITU, the results show generally positive ratings, with median scores of 4, 4, and 3.5 across the experiments. These values, all above the neutral midpoint, suggest that participants showed a moderate to high intention to use the tool in the future.

%Concerning the cognitive ability of users, they all show a good WM capacity, all above the value of 0.8.

\textbf{Responses to RQs.} Table~\ref{tab:hypothesesRej} summarises the results for the hypotheses addressing each RQs. Dark blue cells in the \textit{Hyp.} column indicate NULL hypotheses that have been rejected ({\upshape p}-value $<$ 0.05) for all the three experiments. While light blue cells in the \textit{Hyp.} column indicate hypotheses that have been rejected for two out of three experiments. The table report both significant p-values ($<$  0.05) and weakly significant p-values ($<$ 0.1).
\textcolor{black}{It should be noted that, while all the p-values have been corrected through the Holm-Bonferroni correction, Table~\ref{tab:hypothesesRej} present adjusted p-values to facilitate interpretation (as highlighted in Section~\ref{sec:dataAnalysisProcedure}). This is done to compare the results with the ``classic'' significance level (i.e., 0.05) without having a different reference value for each dependent test. }

\begin{table*}[t]
    \renewcommand{\arraystretch}{1.2}
    \caption{Statistics summary.\\{\scriptsize Dark blue cells in \textit{\textbf{Hyp.}} column indicate NULL hypotheses that have been rejected ({\upshape p}-value $<$ 0.05) for all the experiments.\\
    Light blue cells in \textit{\textbf{Hyp.}} column indicate NULL hypotheses that have been rejected for two out of three experiments.\\
    Dark yellow cells in \textit{\textbf{Effect-size}} column indicate large effect size (value $>$ 0.8).\\
    Light yellow cells in \textit{\textbf{Effect-size}} column indicate medium effect size (value $>$ 0.5).\\
    * indicates that results are weakly significant ({\upshape p}-value $<$ 0.1).\\
    ** indicates that results are significant ({\upshape p}-value $<$ 0.05).}}
    \label{tab:hypothesesRej}
    %\smallskip
    \centering
    \scriptsize
    \vspace*{-0.05cm}
    \begin{tabular}{|c | c | c| c|c |c |c |c|c| |c|c|}
    \hline 
       &  &   & \multicolumn{2}{c|}{\bf{\em Original Experiment}} &  \multicolumn{2}{c|}{\bf{\em First Replication}} & \multicolumn{2}{c||}{\bf{\em Second Replication}} & \multicolumn{2}{c|}{\bf{\em Meta-Analysis}} \\ \hline 
      {\bf{\em RQs}}& {\bf{\em Hyp.}} & {\bf{\em Variable}}	& {\bf{\em p-value}} & {\bf{\em Effect-size}}	& {\bf{\em p-value}} & {\bf{\em Effect-size}} & {\bf{\em p-value}} & {\bf{\em Effect-size}} & {\bf{\em Estimate}} & {\bf{\em  CI (99\%)}}\\ \hline

\multirow{4}{*}{RQ1} & H$1_0$ & \textcolor{black}{Total} Comprehensibility & \cellcolor{gray!25}0.0449** &	0.4979309	& \cellcolor{gray!8}0.0557*  &	\cellcolor{yellow!65}0.9181952	& \cellcolor{gray!8}0.0817* 	& 0.463572 & \cellcolor{gray!25}0.76 &  \cellcolor{gray!25}[0.57; 0.88]\\      
        & \cellcolor{blue!15}H$2_0$ & Learnability &	\cellcolor{gray!25}\cellcolor{gray!25}0.0449** &	0.48342&	\cellcolor{gray!8}0.0557* &	\cellcolor{yellow!65}0.926013&	\cellcolor{gray!25}0.0125**	& \cellcolor{yellow!25}0.6227992   & \cellcolor{gray!25}0.84 & \cellcolor{gray!25}[0.66; 0.93]  \\   
        & H$3_0$ & Understandability &	\cellcolor{gray!25}0.0449** &	0.3550358&	\cellcolor{gray!8}0.0848*  &	0.4659435&	\cellcolor{gray!8}0.0817* 	& 0.288169 & \cellcolor{gray!25}0.82 & \cellcolor{gray!25}[0.64; 0.92]\\
        & H$4_0$ & Evolving & - & - & - & - & 0.486 &	0.01047585 & - & -\\      
      \hline

\multirow{3}{*}{RQ2} & H$5_0$ &  PEOU &  \cellcolor{gray!25}0.0388** & \cellcolor{yellow!25}0.51 &	0.3996 &	0.14 &	0.1652	& 0.25 & 0.48 & [0.31; 0.66]]\\      
        & \cellcolor{blue!45}H$6_0$ & PU &	\cellcolor{gray!25}3.66e-05**  &	\cellcolor{yellow!65}1 & \cellcolor{gray!25}0.0212**  & \cellcolor{yellow!65}1 & \cellcolor{gray!25}1.40e-04**  &	\cellcolor{yellow!65}1  & \cellcolor{gray!25}0.92 & \cellcolor{gray!25}[0.69; 0.99]\\   
        & \cellcolor{blue!15}H$7_0$ & ITU & \cellcolor{gray!25}5.75e-04**  &	\cellcolor{yellow!65}0.96 &	\cellcolor{gray!8}0.0789*  &	\cellcolor{yellow!25}0.72 &	
        \cellcolor{gray!25}0.0026**  &	\cellcolor{yellow!65}0.81	& 0.62 & [0.44; 0.78]\\    
      \hline\hline
      
{\bf{\em RQs}} & {\bf{\em Hyp.}} & {\bf{\em Relation between variables}}  &   {\bf{\em p-value}}  & {\bf{\em $\rho$ }} &   {\bf{\em p-value}} & {\bf{\em $\rho$ }} &   {\bf{\em p-value}} & {\bf{\em $\rho$  }} & {\bf{\em p-value}} & {\bf{\em  CI (95\%)}}  \\ \hline

\multirow{3}{*}{RQ3} & H$8_0$ &  PEOU $\leftrightarrow$ PU 	& \cellcolor{gray!25}0.0255**  &  $0.5354501$ &	0.1639 	& 0.6484968  	& 0.7366 &	0.07805842   & 0.052 & [-0.004; 0.683]\\      
        & H$9_0$ & PU $\leftrightarrow$ ITU &	\cellcolor{gray!25}0.0255** &	0.5585299 
 & 0.3229		& 0.402469  & \cellcolor{gray!8}0.0522*	&	0.5129636   & \cellcolor{gray!25}2e-04 & \cellcolor{gray!25}[0.265; 0.708]\\  
        
        & \cellcolor{blue!15}H$10_0$ & PEOU $\leftrightarrow$ ITU 	& \cellcolor{gray!25}0.0348**	& 0.4624643   & \cellcolor{gray!25}0.0109** & 0.8834522 		& 0.4966	& 0.2636066 	& \cellcolor{gray!25}0.0277 & \cellcolor{gray!25}[0.067; 0.82] \\
      \hline

\multirow{12}{*}{RQ4} & H$11_0$ &  \textcolor{black}{Total} Comprehensibility $\leftrightarrow$ PEOU & 1 &	0.2330876  &	1	& 0.1575787 &	1 &	0.08550979    & 0.3018 & [-0.144; 0.436] \\      
        & H$12_0$ & \textcolor{black}{Total} Comprehensibility $\leftrightarrow$ PU &	 1 &	0.03873335 & 1	 & 0.2349398	 &	1	& 0.1355782 & 0.4969 & [-0.197;0.39]\\   
        & H$13_0$ & \textcolor{black}{Total} Comprehensibility $\leftrightarrow$ ITU & 1 &	0.08753551 &	1	& 0.03658809 &	1	& 0.1593746	& 0.4671 & [-0.19; 0.397]\\
        & H$14_0$ & Learnability $\leftrightarrow$ PEOU & 1 &	0.05060749 & 1		& 0.06260879 &	1 & -0.1102078 & 0.9047 & [-0.314; 0.28]	\\    
        & H$15_0$ & Learnability $\leftrightarrow$ PU & 1  &	-0.2057128	& 1	& 0.5116729 & 1	 &	0.1600668 & 0.7228 & [-0.287; 0.402]\\
        
        & H$16_0$ & Learnability $\leftrightarrow$ ITU & 1  &	-0.09011384 &	1	& -0.1259882 & 0.5653	 & 0.437877	& 0.5407 & [-0.284; 0.505]
        \\
        & H$17_0$ & Understandability $\leftrightarrow$ PEOU & 1 &	0.2072203 &	1 & 0.2533473	 &	1	& 0.2352589 & 0.1423 & [-0.077; 0.489]\\ 
        & H$18_0$ & Understandability $\leftrightarrow$ PU & 1 & 0.1310361	 &	1 & -0.02518165	 &	1	& 0.3162398 & 0.2036 & [-0.107; 0.466]\\ 
        & H$19_0$ & Understandability $\leftrightarrow$ ITU & 1 & 0.04423699	 &	1 & 0.2294157	 &	1	& 0.1411804 & 0.48 & [-0.193; 0.394]\\ 
        & H$20_0$ & Evolving $\leftrightarrow$ PEOU & - &	-&	-&	-&	1 &	0.1170276 & - & -\\
        & H$21_0$ & Evolving $\leftrightarrow$ PU & -&	-	&-	&-&	1	& -0.169169 & - & -\\
        & H$22_0$ & Evolving $\leftrightarrow$ ITU & -&	-	&-&	-&	1 & -0.161396	& - & -\\
      \hline

%\multirow{4}{*}{RQ5} & H$23_0$ &  WM Capacity $\leftrightarrow$ \textcolor{black}{Total} Comprehensibility & 0.3208  & 0.3177995	& 1	& -0.15569148
%	&	0.5524	&  -0.3346329 \\      
%        & H$24_0$ & WM Capacity $\leftrightarrow$ Learnability & 0.8085   & 0.05628297	& 0.2012	& -0.6735753	&	0.6866	&  -0.2108071 \\   
%        & H$25_0$ & WM Capacity $\leftrightarrow$ Understandability &  \cellcolor{gray!25}0.0282** & 0.5524079	& 1	& 0.20025059	& 0.5524		& -0.3044096 	\\
%        & H$26_0$ & WM Capacity $\leftrightarrow$ Evolving & -&	-	&-	&-	& 0.6866 & -0.217634411	\\    
%        \hline

   \end{tabular}
\end{table*}

\medskip

\noindent\textbf{RQ1. }
The results reported in Table~\ref{tab:hypothesesRej} indicate that learnability is the only dimension
exceeding the target value of 0.6 in two out of three experiments, thus allowing rejection of  H$2_0$, with statistically significant results ($p < 0.05$) in both the original experiment and the second replication. 
In the first replication, learnability also reaches weak significance ($p = 0.557$).
Total comprehensibility and understandability reach at least weak significance in all three experiments: both are strongly significant in the original study and weakly significant in the two replications. These consistent results across all experiments suggest a trend toward rejecting H$1_0$ and H$3_0$.
\textbf{These findings suggest a sufficient overall comprehensibility of the Neverlang meta-language, particularly regarding its syntax.}

The evolving dimension (H$_{40}$) was assessed only in the second replication and did not yield significant results ($p = 0.486$). While this can suggest that the comprehensibility ``in action'', i.e., when users are requested to modify a Neverlang program, is not sufficient, no final conclusion can be drawn, since we cannot reject the NULL hypothesis. Modifying a Neverlang program is arguably the hardest part of the test, and, given the limited training the participants received, it is understandable that they underperformed when asked to use the Neverlang meta-language. 

The effect sizes for learnability range from moderate to large, with the highest observed in the first replication. This might be attributed to the more structured and extended training delivered in that iteration or to differences in participant backgrounds.
The weak significance levels observed in the first replication for all three hypotheses (H${10}$, H${20}$, and H$_{30}$) could also be explained by its relatively small sample size, which may have reduced the statistical power of the analysis.

\medskip
\noindent\textbf{RQ2. }
\textcolor{black}{The test results for the acceptance variables (i.e., PEOU, PU, and ITU) reveal that PU exhibits significant values surpassing the neutral value of 3 across all three experiments, while ITU shows significant results in the original experiment and the second replication, with weakly significant results in the first replication (p-value $<$ 0.1) (cf.\ Table~\ref{tab:hypothesesRej}). \textbf{These findings indicate a favourable perception of the tool's usefulness and a general intention to utilise it in the future, albeit with some variability across experiments.}}

Regarding PEOU, only the results from the original experiment are statistically significant, with a p-value of 0.0388 and a medium effect size. \textbf{This suggests that the language requires improvements to facilitate its ease of use}.

\medskip
\noindent\textbf{RQ3.}
The relationships among acceptance variables show varying levels of robustness across experiments.
The relationship between perceived ease of use (PEOU) and perceived usefulness (PU) (H$_{8_0}$) is statistically significant only in the original experiment.

In contrast, stronger empirical support is observed for the relationships between PEOU and intention to use (ITU), and PU and ITU.
In the first case (PEOU $\leftrightarrow$ ITU), results are significant ($p < 0.05$) in two out of three experiments.
In the second case (PU $\leftrightarrow$ ITU), the relationship is statistically significant in the original experiment ($p < 0.05$), and approaches conventional significance thresholds in the second replication ($p = 0.0522$), indicating a consistent trend across studies.

\textbf{This suggests that the intention to use Neverlang (or not using it), is associated with both perceived ease of use and perceived usefulness.}

\medskip
\noindent\textbf{RQ4. } 
As depicted in Table~\ref{tab:hypothesesRej}, none of the NULL hypotheses linked with RQ4 have been rejected. 
\textbf{These outcomes suggest that there is no discernible relationship between the comprehensibility of the Neverlang meta-language and user acceptance of the tool}.

\subsection{Results: Family Data Analysis}
\begin{figure}[t]
    \centering

    \begin{subfigure}[t]{0.47\columnwidth}
        \centering
        \includegraphics[width=\linewidth]{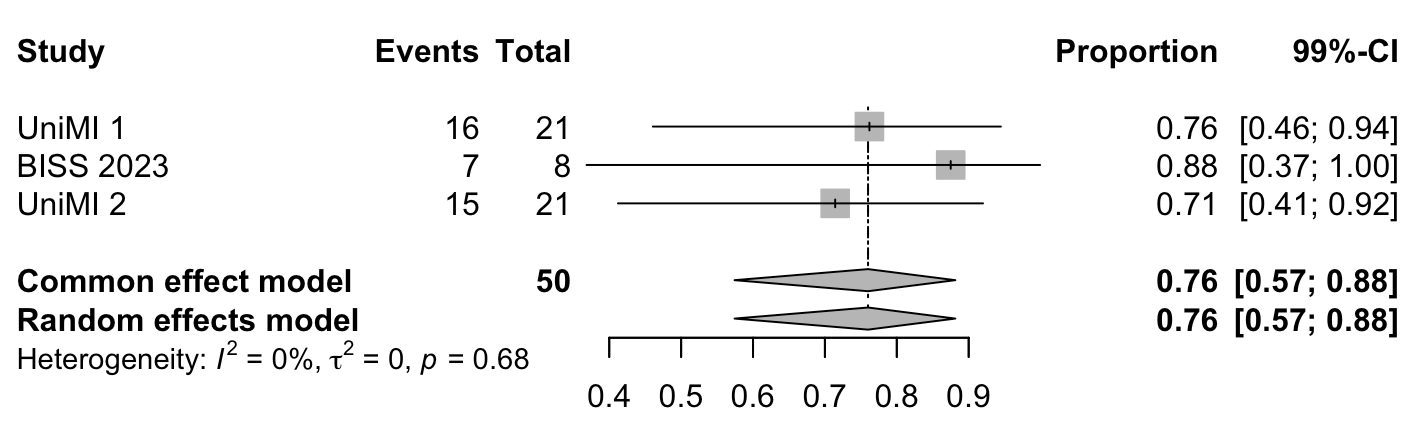}
        \caption{Total Comprehensibility ≥ 0.6.}
        \label{fig:forestPlot}
    \end{subfigure}
    \hfill
    \begin{subfigure}[t]{0.47\columnwidth}
        \centering
        \includegraphics[width=\linewidth]{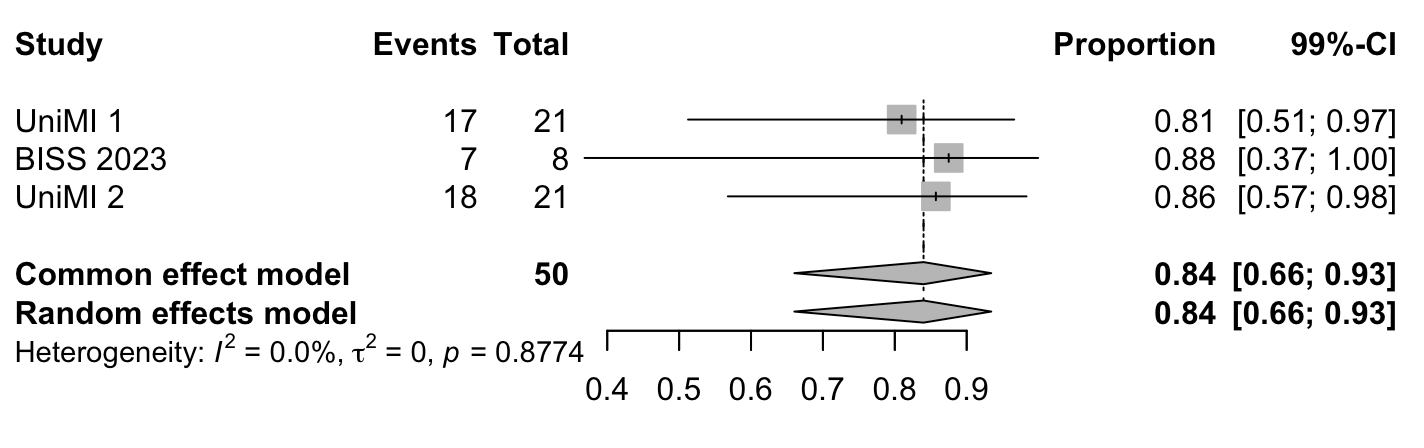}
        \caption{Learnability ≥ 0.6.}
        \label{fig:forestPlot2}
    \end{subfigure}

    \vspace{1.5em}

    \begin{subfigure}[t]{0.47\columnwidth}
        \centering
        \includegraphics[width=\linewidth]{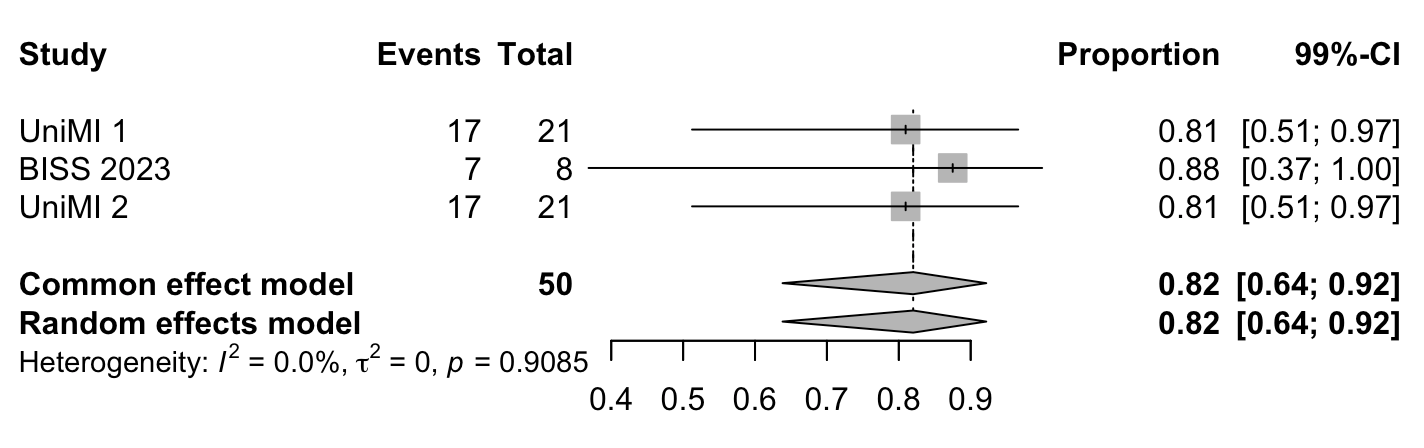}
        \caption{Understandability > ≥ 0.6.}
        \label{fig:forestPlot3}
    \end{subfigure}
    \hfill
    \begin{subfigure}[t]{0.47\columnwidth}
        \centering
        \includegraphics[width=\linewidth]{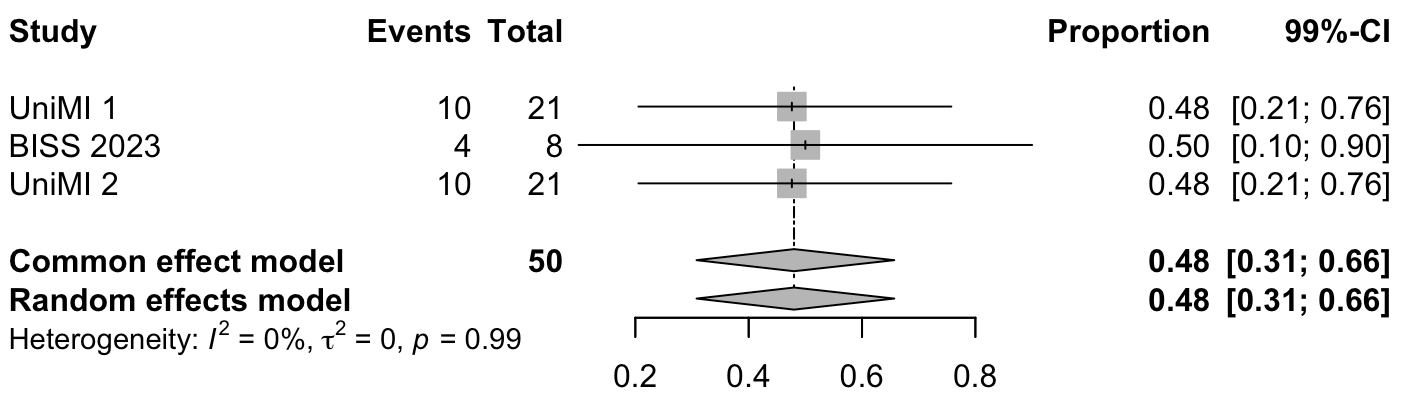}
        \caption{Perceived ease of use > 3.}
        \label{fig:forestPlot4}
    \end{subfigure}

    \vspace{1.5em}

    \begin{subfigure}[t]{0.47\columnwidth}
        \centering
        \includegraphics[width=\linewidth]{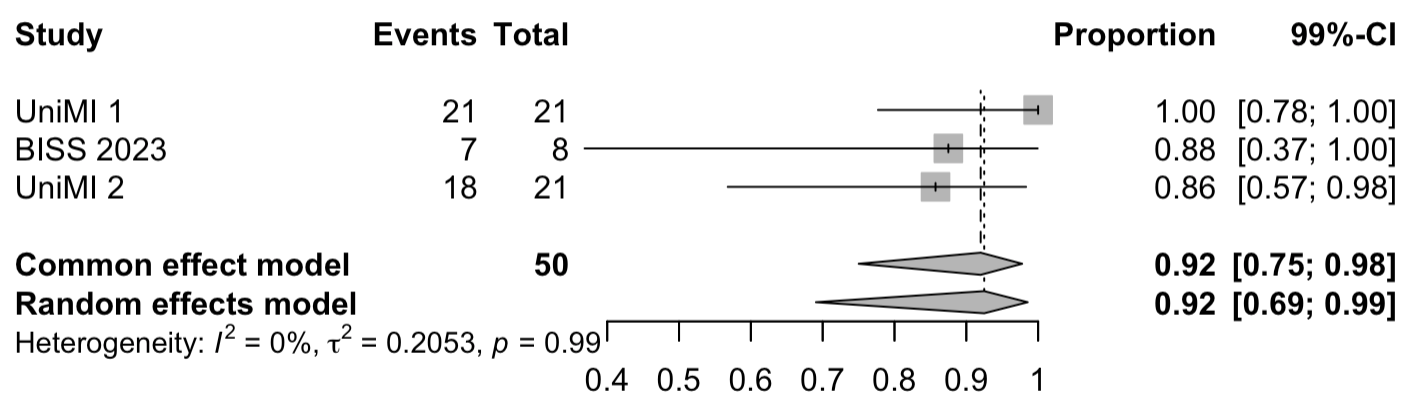}
        \caption{Perceived usefulness > 3.}
        \label{fig:forestPlot5}
    \end{subfigure}
    \hfill
    \begin{subfigure}[t]{0.47\columnwidth}
        \centering
        \includegraphics[width=\linewidth]{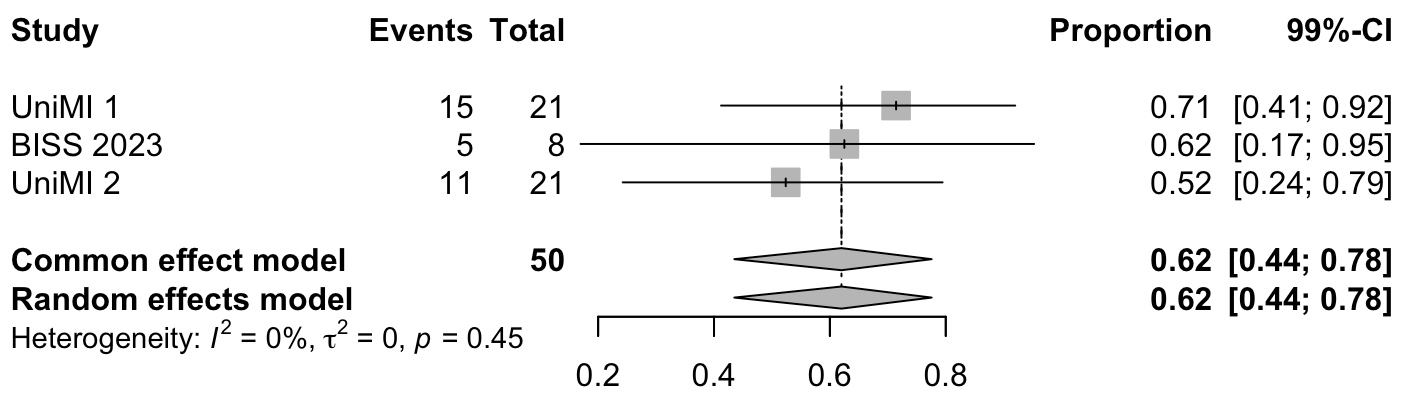}
        \caption{Intention to use > 3.}
        \label{fig:forestPlot6}
    \end{subfigure}

    \caption{Forest plots summarizing the results of the meta-analyses of proportions across the three experiments for each measured dimension.}
    \label{fig:forestPlotsAll}
\end{figure}

The results of the Meta-Analysis show overall consistency across the experiments and confirm some of the findings from previous tests (cf. Table~\ref{tab:hypothesesRej}).
For what concerns RQ1 (cf. Figures \ref{fig:forestPlot}, \ref{fig:forestPlot2}, and \ref{fig:forestPlot3}), the results of the meta-analysis of proportion confirm that the syntax (i.e., learnability) is the most understood dimension of the Neverlang meta-language, with an overall pooled proportion of users exceeding the 0.6 threshold of 0.84 (CI = [0.66; 0.93]). This align with the hypothesis testing results, where learnability showed statistically significant results in two out of three experiments and weakly significant result (approaching significance) in the third experiment. 
Results for total comprehensibility and understandability also broadly support the outcomes of the hypothesis testing. The pooled proportion of users exceeding the 0.6 threshold is 0.76 for total comprehensibility (99\% CI = [0.5744; 0.8814]) and 0.82 for understandability (99\% CI = [0.64; 0.92]). In comparison, the corresponding hypothesis tests showed strong significance in the original experiment and weak significance in the two replications.

Since none of the confidence intervals include 0.5, all three meta-analytic results are considered statistically significant. Heterogeneity analysis revealed low levels of heterogeneity, with $I^2 = 0.0\%$. This suggests that the variability in proportions among the studies can be attributed to random variation rather than true differences in effect sizes.
Although H$1_0$, H$2_0$, and H$3_0$ were not formally rejected in every experiment, the convergence of findings across studies and the meta-analytic estimates consistently exceeding the threshold of interest suggest a coherent and positive pattern in users’ comprehension of the Neverlang meta-language.

Regarding acceptance (RQ2), we analyse the three variables (i.e., PEOU, PU, and ITU) individually. The results confirm that ease of use is the least favourably perceived aspect among the acceptance variables. As Fig.~\ref{fig:forestPlot4} shows, the overall proportion of users that perceive Neverlang as easy to use is $0.48$, with a $99\%$ confidence interval of [0.31; 0.66]. Since this interval includes $0.5$, the result is not statistically conclusive. While the point estimate suggests that fewer than half of the users found the tool easy to use, the confidence interval reflects uncertainty about this conclusion. The heterogeneity analysis indicates no variability among the three experiments ($I^2 = 0.0\%$), with proportions of 0.48 in both the original experiment (UniMi1 2022) and the second replication (UniMi~2 2023), and 0.50 in the first replication (BISS 2023).

The results on intention to use (cf.\ Fig.~\ref{fig:forestPlot6}) show higher values, with an overall proportion of $0.62$ and a $99\%$ confidence interval of [0.44; 0.78]. Although the point estimate is above 0.5 (indicating that more than 50\% of participants rated ITU positively), the confidence interval includes $0.5$, indicating that the result is not statistically significant at the $95\%$ level. This suggests a positive trend in users’ intention to use Neverlang, but further evidence is needed to confirm this conclusively. The heterogeneity analysis indicates no significant variability among the three experiments, suggesting that the differences in proportions---$0.71$ in the original experiment (UniMi~1 2022), $0.62$ in the first replication (BISS 2023), and $0.52$ in the second replication (UniMi~2 2023)---are not due to true differences in effect size, but rather attributable to other confounding factors.

The results on perceived usefulness show the highest proportion (cf.\ Fig.~\ref{fig:forestPlot5}). The proportion of users rating the tool as useful was $100\%$ in the original experiment (UniMi~1 2022), $87.5\%$ in the first replication (BISS 2023), and $85.7\%$ in the second replication (UniMi~2 2023). The random effects model estimates an overall perceived usefulness proportion of $0.9249$ with a $95\%$ confidence interval of [0.6906; 0.9855]. This result is statistically significant and indicates a strong and consistent perception of usefulness across the studies. The test for heterogeneity shows some degree of variability in perceived usefulness among the studies ($\tau^2 = 0.2053$), but based on the $I^2 = 0.0\%$ and a non-significant $p$-value ($p = 0.99$), this variability is not substantial enough to indicate meaningful differences between the studies.

These meta-analytic findings partially align with the hypothesis testing results. PU showed statistically significant results across all experiments and a significant meta-analytic proportion estimate, reinforcing the overall positive perception of the tool’s usefulness. For ITU, while hypothesis testing indicated significant or weakly significant results across the three experiments, the meta-analysis did not yield statistically conclusive evidence, as the confidence interval included $0.5$. This discrepancy highlights that, despite promising indications in individual studies, further evidence is needed to confirm a strong overall intention to adopt the tool. Lastly, for PEOU, both the meta-analysis and hypothesis testing converge in showing weaker support, suggesting that ease of use remains the main challenge.

For what concerns the relationship between the acceptance variables (RQ3), the meta-analytic results show statistically significant relationships between both PU and ITU, and PEOU and ITU. Specifically,
The overall correlation between PU and ITU was strong and significant ($\rho = 0.521$, 95\% CI = [0.265; 0.708], $p = 0.0002$), as was the relationship between PEOU and ITU ($\rho = 0.545$, 95\% CI = [0.067; 0.820], $p = 0.0277$). Although the correlation between PEOU and PU approaches statistical significance ($p = 0.052$), the 95\% confidence interval includes zero (CI: [–0.004; 0.68]), suggesting that the association is not statistically reliable.
These meta-analytic findings align with the outcomes of the hypothesis tests performed on the individual experiments. The relationship between PEOU and ITU was significant in two out of three experiments, while the relationship between PU and ITU was significant in the original experiment and approached significance in the second replication. Conversely, the relationship between PEOU and PU was only significant in the original experiment and not replicated in the other two. These results reinforce the importance of both perceived ease of use and perceived usefulness in shaping users’ intention to use Neverlang, but suggest a weaker and less consistent link between PEOU and PU. 
Finally, regarding the relationship between acceptance variables and comprehensibility dimensions, the meta-analysis of the correlation coefficients did not produce any statistically significant results. All correlations had $p$-values well above 0.1, and their confidence intervals included zero. 
These findings are consistent with the hypothesis testing results, where none of the corresponding null hypotheses were rejected across the three experiments. This suggests that acceptance and comprehensibility are not linearly related in this context: users may appreciate or intend to use Neverlang regardless of their actual level of comprehension, and vice versa.

\section{Discussion}\label{sect:discussion}
The results of the family of experiments revealed remarkable consistency in both the rejection and failure of rejection of the hypotheses, strengthening the reliability of our findings. Moreover, the results of the meta-analysis further enrich our findings. %about the comprehensibility and acceptance of Neverlang, showing the common proportion of users whose comprehensibility is above sufficient evaluation and whose PEOU, PU, and ITU are above neutral perception.

RQ1, RQ2, and RQ3 received, to some extent, affirmative responses through our analyses.

Regarding \textbf{RQ1}, which investigates the comprehensibility of the Neverlang meta-language both globally and across specific dimensions, the results reveal a coherent and encouraging pattern across the three experiments. Although statistical significance levels varied---being strong in the original experiment, weak in the first replication, and mixed in the second---the overall trend suggests that users were generally able to understand Neverlang meta-language to a sufficient degree. 
The weak significance in the first replication for all the hypotheses (i.e., H$1_0$–H$4_0$) could be attributed to the smaller sample size in that experiment (8 participants compared to 21 in both the original and second replication).
Notably, learnability, which reflects users’ ability to understand Neverlang meta-language’s syntax, emerged as the most consistently positive dimension, showing significant or near-significant results in all replications.  

These results are further supported by the meta-analysis of proportions, which confirms the overall comprehensibility of the meta-language, especially at the syntactic level. While not all hypotheses (H$1_0$–H$3_0$) were formally rejected in every individual experiment, the aggregated results point toward a reliable pattern of sufficient user comprehension. In particular, the meta-analysis shows that more than two-thirds of users reached a sufficient level of evaluation not only globally (i.e., total comprehensibility), but also for both the syntactic (learnability) and semantic (understandability) dimensions. This reinforces the view that Neverlang’s syntax is the most accessible aspect of the language, as also indicated by learnability being the only dimension that achieved statistically significant results in two out of three experiments.
Nonetheless, the narrow confidence intervals for total comprehensibility, combined with the absence of heterogeneity ($I^2 = 0\%$), suggest that this represents a reliable population-level estimate across varying training conditions and user backgrounds. While these results confirm Neverlang’s overall comprehensibility, the higher performance observed in the first replication (88\%), which included a longer training activity and involved Ph.D.\ students, suggests that targeted training interventions could further increase comprehension rates, and that users with a higher education profile may perform better. Additional studies are needed to explore these hypotheses.

In contrast, evolving---the dimension related to users’ ability to modify Neverlang programs---did not yield statistically significant results in hypothesis testing (H$4_0$) and was assessed only in the second replication. As a result, it could not be included in the meta-analysis, which limits the generalisability of this finding and highlights the need for further investigation into users’ ability to apply the language in practice.

Taken together, the findings suggest that Neverlang’s syntax is generally accessible to users, who can grasp its fundamental constructs effectively.
However, the results on the other dimensions suggest that there are still intrinsic complexities associated with Neverlang's meta-language. We hypothesise that modularity may be one of the barriers to comprehending  Neverlang. Indeed, the learnability questions---which received higher scores on the test---did not require a deep understanding of this language feature, which was instead highly prominent in understandability and evolving questions. This is also confirmed by the authors' experience in teaching Neverlang. The modular approach, which involves small, discrete slices of code, can be challenging for those accustomed to writing large, monolithic blocks of code. This difficulty is compounded by the necessity to understand advanced concepts like grammar semantics and the visitor pattern, which may not be familiar to the participants. Further studies are required to confirm this hypothesis.

Regarding \textbf{RQ2}, which assesses the acceptance of Neverlang by users in terms of PEOU, PU, and ITU, the last two variables received good scores, indicating that the participants understood the practical potential of the language. Instead, PEOU emerged as the most negatively perceived variable, with significance levels higher than $0.05$ in both the first and the second replication. In this regard, while the tool received general acceptance, ease of use remained a concern. This could, in principle, be due to the short video training received in the original experiment and second replication. However, we can rule out this hypothesis, as the participants of the first replication, who received 12 hours of training, also had similar opinions. Again, modularity can play a role, and thus more hands-on training targeting this feature may be necessary. It should also be recalled that Neverlang is an academic product, like many other workbenches, and may lack sufficient documentation and libraries, which could facilitate its usage. This is confirmed by the feedback of one of the participants: ``[\ldots] I would like a more advanced comment system, to provide access to documentation''.   
Enhancing the documentation and providing comprehensive libraries of reusable components could improve ease of use, and, more generally, user acceptance. 

These findings are aligned with the results of the meta-analysis of proportions. While PU showed a high overall proportion of positive responses and was statistically significant, with on average more than 90\% of users perceiving Neverlang as useful, ITU received a lower proportion (0.62) and was not statistically conclusive (CI = [0.44; 0.78]). The results for PEOU were even less favourable, with a pooled proportion of 0.48 (CI = [0.31; 0.66]), clearly confirming that ease of use is the most critical issue in user acceptance.

Concerning \textbf{RQ3}, we have observed that both PEOU and PU are associated with ITU in the majority of the experiments, while the association between PEOU and PU was observed only in the original one. This partially confirms the core structure of the MEM reference model, in which usefulness and ease of use are assumed to be associated with the intention to use a tool or method. This suggested that Neverlang is not substantially different from other technologies, for which this relationship was observed (cf., e.g.,~\cite{abrahao2011evaluating}). On the other hand, the non-confirmed relation between PEOU and PU suggests that possible difficulties with the tools do not affect its perceived usefulness. This divergence from the MEM model’s original framework may stem from contextual factors specific to Neverlang’s domain. Unlike general-purpose technologies where ease of use often amplifies perceived usefulness (e.g., consumer software), Neverlang---as a language workbench tailored for modular language development---might cater to users who prioritise functional capabilities over usability. Users may dissociate ease of use from usefulness, particularly if the tool’s value lies in its technical expressiveness or adaptability, even at the cost of complexity. This means that users might tolerate steeper learning curves if the tool enables unique functionalities. These findings highlight the importance of contextualising technology acceptance models. While MEM provides a foundational framework, its assumptions must be adapted to reflect domain-specific user priorities and workflows. For tools like Neverlang, this means balancing usability enhancements with clear communication of technical advantages to align with the expectations of specialised developer communities.

These results are again supported by the meta-analysis of correlation coefficients. While in hypothesis testing only H$10_0$ (PEOU $\leftrightarrow$ ITU) was rejected in two out of three experiments, the relationship between PU and ITU (H$9_0$) was significant in the original experiment and weakly significant in the second replication (approaching the conventional threshold at $p = 0.0522$). In the meta-analysis, both these relationships (H$10_0$ and H$9_0$) were found to be statistically significant, confirming a consistent association between perceived ease of use, perceived usefulness, and intention to use across the studies. This further strengthens the conclusion that user intention to use Neverlang is primarily driven by its perceived ease of use and, to a slightly lesser extent, by its usefulness.

Concerning \textbf{RQ4}, the relationships between variables were non-significant. These results were consistent across all three experiments, suggesting that actual success in using Neverlang (comprehensibility) does not influence perceived success (acceptance) regarding the tool. 

These results may be mainly due to the low PEOU that does not align with the high scores obtained in the comprehension test. Essentially, those who performed better in the comprehension test did not do so due to the language's simplicity, but for other reasons, e.g., a greater predisposition to programming, higher abilities in related tasks, or greater attention. In addition, we note that in some cases there is even a negative correlation---although not significant between comprehensibility and acceptance variables. This could be influenced by the fact that a deeper comprehension of the language, i.e., higher success in the test, enables users to better identify its critical issues, thus leading them to judge it more negatively. Understanding the critical issues should be explored through qualitative studies in which more proficient users are interviewed to understand perceived strong and weak points. 

These non-significant results were confirmed by the meta-analysis of correlation coefficients, in which none of the hypotheses yielded statistically significant associations.

\medskip
\noindent\textbf{Implications for Research. } Our results underscore several key areas for future research on Neverlang and language workbenches in general. First, comparative studies with other language workbenches are essential to understand the unique challenges and advantages of each tool. This comparative approach, supplemented by qualitative user feedback, could help identify best practices and common issues across different tools, potentially leading to the establishment of field standards. 

Such comparisons could also be conducted through the development of DSLs to evaluate the strengths and weaknesses of various tools, as done in previous judgement studies~\cite{MBBF20} and systematic evaluations~\cite{MBB22} targeting formal languages, which arguably share some of the hurdles of workbench languages in terms of ease of use. However, finding experts proficient in multiple tools may present a significant challenge. 

Longitudinal studies are also needed to assess the long-term usability and re-usability of Neverlang modules, which is one of its main advantages. Evaluating how well users can recognise and utilise reusable modules will provide valuable insights into the language's practicality and efficiency.

The main point of improvement that emerged from the study is the need to increase the ease of use of the meta-language. The limited perceived ease of use, compared to the other acceptance variables, could be due to the misalignment between the syntax of Neverlang and the one of classical GPLs, which were more familiar to the participants. To understand if this is the source of the problem, an interesting avenue of research could be asking participants to directly use the Java language underlying Neverlang\footnote{Neverlang is a DSL built on top of Java to ease the development of other DSLs. However, Java could be directly used as an alternative.}. While this would require more instructions, participants could grasp the concepts and constructs more effectively. On the other hand, to better study potential solutions to improve ease of use, researchers should consider using more user-friendly GUIs for Neverlang, such as AiDE, which is tailored for Language Product Lines~\cite{Cazzola20}. Although this would not made available all the advanced capabilities of the meta-language, novice users may find it more suitable to gain confidence with the technology.

\medskip
\noindent\textbf{Implications for Teaching. }
At this stage, Neverlang can be primarily regarded as an educational tool, especially for teaching modular and declarative programming principles. However, to maximise its effectiveness, it needs to become more accessible and user-friendly. This involves the identification of the specific constructs and concepts that are hard for students to digest and the development of targeted strategies to teach them. Facilitating student access to the language also entails creating tutorials with practical examples and exercises. 
It should be noted that both the video tutorial and the 12-hours course were mainly traditional lectures, rather than a learning-by-doing training. 
A long-term, incremental course structure with hands-on activities is necessary for students to not only learn Neverlang but to also appreciate its potential and strengths. This approach should include training on how to easily create DSLs tailored to specific fields, which can be readily used by non-IT experts.  Additionally, an incremental learning approach could reduce the barriers to accessing the advanced features of the tool, and thus better exploit its potential.

\medskip
\noindent\textbf{Implications for Practice. }
From a practical perspective, our findings show that participants have a clear understanding of the usefulness of the language and exhibit a willingness to adopt it if needed. This suggests that introducing Neverlang in a real-world environment is feasible, and, as long as concerns about ease of use are addressed, there are no significant barriers to acceptance. However, since Neverlang is an academic tool, there are still several improvements to introduce to make it industry-ready. In particular, developing libraries of language components and functions similar to those found in traditional programming languages is a priority. Supporting frameworks and powerful integrated development environments (IDE) specialised for the creation of DSLs are other assets that are currently missing. 

Neverlang's greatest strength is its modularity; however, this is also the most complex aspect of the language. As a result, its potential success does not automatically translate into greater ease of use. Our findings clarify this intrinsic trade-off between Neverlang’s powerful capabilities and its usability. As a possible improvement, it would be beneficial to offer different interfaces that are tailored to the types of users, making the language adaptable to users with varying skill levels. By enabling the access to different levels of complexity and power offered by the meta-language, this could become more accessible while retaining its powerful features.

\section{Threats to Validity} \label{sect:threats}
%We address the threats to validity following the guidelines provided by Wohlin et al. \cite{wohlin2012experimentation}.
%, focusing specifically on construct validity, internal validity, external validity, and conclusion validity.

\subsection{Construct Validity}
To assess comprehensibility, we calculated the average of correct answers divided by the total number of questions, a well-established technique in the literature~\cite{abrahao2011evaluating,moody2001dealing,broccia2024assessing}. The comprehensibility dimensions (i.e., learnability, understandability, and evolving) are widely employed in the evaluation of languages~\cite{kosar2012program,kosar2010comparing,kosar2009influence}.  They were adapted for the evaluation of Neverlang from the template provided by Kosar et al.~\cite{kosar2012program}, revised by two Neverlang experts, and deemed appropriate for assessing the comprehensibility of the Neverlang meta-language.
The evaluation of the evolving dimension in the second replication was performed manually, so subjectivity could have played a role. However, the involvement of two Neverlang experts in the assessment mitigates this threat. More structured and longitudinal tasks (e.g., composing modules and slices to create new DSLs) were not considered in this study, and different outcomes might be observed with different tasks.

\textcolor{black}{A threshold of 0.6 was selected as an indicator of a sufficient degree of comprehension by the participants. The choice of different thresholds could affect the results. However, the questions used to measure comprehensibility were defined with the support of two Neverlang experts, which ensures content validity. These questions were carefully crafted so that correctly answering 60\% of them would indicate sufficient comprehension according to the judgement of the experts. To assess the reliability of the comprehensibility questionnaire, we computed the McDonald's Omega, which resulted in a value of 0.7, indicating an acceptable reliability~\cite{mcneish2018thanks}. }

Users' acceptance was assessed through existing models~\cite{moody2001dealing}, and the questionnaire was tailored to the Neverlang evaluation according to previous studies~\cite{abrahao2011evaluating,broccia2024assessing}. 

To measure WM capacity, we used two WMSTs~\cite{conway2005working}, which are widely used tasks in cognitive psychology and in the field of empirical software engineering~\cite{mansoor2023empirical}.

A further construct validity threat is introduced by the self-assessment of the Post-study phase, when we collected participants' prior knowledge. The assessment was based on a Likert scale, and we specified the judgements associated with each score on the scale (``very low'', ``low'', ``medium'', ``high'', ``very high''). An alternative could have been to ask about their experience in terms of years. However, since participants were students and had similar numbers of years of experience across the different dimensions (programming skills, Java and object-oriented programming skills, and experience with language workbenches), we arguably considered self-assessment a more effective way to characterise them. We also note that, given the opportunistic sampling, we did not attempt to connect participants' experience, and demographics in general, with the experiment outcomes. Thus, a precise characterisation of the participants' background knowledge was not strictly required. 

\subsection{Internal Validity}
We made an effort to maximise internal validity by evaluating different quantitative and objective variables. Subjective opinions were collected using the user acceptance questionnaire, which could have introduced participant bias. \textcolor{black}{To mitigate systematic response bias, we shuffled the questions and mixed positive and negative statements.}

Participants never met the experimenter, as all studies were conducted online. This approach helped limit the potential for participants to provide positive answers to please the researchers, thereby reducing the Hawthorne effect \cite{noland1958landsberger}.  

This paper considers the relationship between WM capacity and comprehensibility. Other confounding factors, like user expertise and background experience, which might affect comprehensibility, are ignored. Discrepancies among experiments possibly associated with background and demographic factors---especially in the first replication---are, however, highlighted in the paper. It is also important to note that this work measures correlation, not causation.

Besides, this study is not designed to compare language workbenches. Therefore, we do not investigate the comprehensibility of Neverlang in comparison to other platforms.

While the original study and the second replication included only a short preliminary training, the first replication included a long introductory seminar. This could have led to different results, which would make the experiments hardly comparable. Other differences in the experiments, highlighted in Section~\ref{sec:differences}, could have also impacted the soundness of the comparison. However, we notice that the results are rather uniform across replications, which suggests that this factor did not affect the outcomes. Minor differences in the results are also highlighted in the results and discussion sections (cf.\ Sections~\ref{sec:results} and~\ref{sect:discussion}). 

\subsection{External Validity} 
Participants were opportunistically chosen from the academic field, varying in seniority. The selected participants encompassed diverse genders, experience levels, and backgrounds, enhancing generalisability. 

While employing students in software engineering experiments generally limits the generalisability of the results because they may not represent the broader population of professional software engineers, 
students can be suitable participants for certain types of experiments, particularly those focused on learning and educational aspects \cite{santos2018analyzing}.
Our family of experiments focused on assessing the comprehensibility, including learnability, of the Neverlang meta-language. In this context, students are arguably the most suitable participants. \textcolor{black}{Additionally, as Neverlang is a novel tool without widespread adoption, focusing on novices enables us to evaluate its comprehensibility and acceptance from a first-time user perspective---ensuring our findings directly apply to new adopters.
It should also be noted that expert users, who are already used to specific tools or languages, are typically not so keen on using new instruments~\cite{venkatesh2008technology}.}

\textcolor{black}{The similarly high WM capacity, which characterises all participants, may have influenced the results, as a more varied range of WM capacity levels could potentially lead to different outcomes. 
This homogeneity may stem from common characteristics of our sample, e.g., the participants' computer science background or young age. Both these characteristics are known to correlate with higher cognitive abilities~\cite{shute1991likely,salthouse2009does}. It should be remarked that a computer science background is a desirable characteristic for our test, as our reference population is computer science students, who are typically young.  %The selection of young participants is conditioned by our target population, i.e., students. 
 %the available population for the test, which primarily consists of university students. 
 %This age group is naturally overrepresented in our sample due to recruitment constraints, as it commonly happens in many studies in software engineering~\cite{falessi2018empirical,salman2015students}. 
 To mitigate the observed homogeneity in terms of WM capacity, more fine-grained tests capable of distinguishing between similar levels of WM capacity may be needed. However, to the best of our knowledge, such tests are not currently available.
 Moreover, identifying and selecting participants with significantly different WM capacity levels a priori was not feasible due to the difficulty of recruiting individuals who had an adequate profile for the task and were available to perform the test. Overall, given our sample, our results can be considered applicable to subjects with high WM. 
 }
%However, this sample may not fully represent all Neverlang users, potentially influencing the study results, and different outcomes may be observed in daily practice with Neverlang. Further research involving users from diverse fields is needed to confirm the applicability of our conclusions across all user classes.

Except for the second replication, which included an evolving dimension, i.e., changing a program, we mainly evaluated Neverlang in a code \textit{reading} task. Different results might be observed if a coding task is considered. It must be noted that this study is a controlled experiment aimed at maximising internal validity and does not evaluate Neverlang users in a realistic setting where contextual factors play a relevant role. Therefore, case studies are needed to confirm that our conclusions apply in real-life environments. 

Experimental mortality may also have biased the results. The mortality was as follows. The original experiment had 24 participants, 21 of whom completed the test; the first replication was 14 vs.~8; the second replication 25 vs.~21. This experimental mortality, although generally acceptable (21\% in total), could affect the generalisability of the results, as those who completed the test could also be the most motivated participants. 

\subsection{Conclusion Validity}
Our subject selection was opportunistic, and, while the size of the sample considered in the original experiment and second replication (21~participants each) is \textcolor{black}{comparable} with typical sample sizes in empirical software engineering---from~10 to 30~subjects~\cite{santos2018analyzing}---the first replication included 8~subjects. \textcolor{black}{It should be noted that for families of experiments the majority of families include between~50 and 100~participants. Our study includes 50~participants, thus falling in this spectrum. Furthermore, according to the mapping study on families of experiments in software engineering \cite{santos2018analyzing} about 50\% of the surveyed families include 3 experiments, like in our case.}

\textcolor{black}{It should be noted that, although the experiments were conducted online, the pool of potential participants was inherently limited due to the time commitment required to complete each experiment. Specifically, completing each experiment involved multiple phases over two separate days, requiring sustained participant engagement. This constraint significantly reduced the feasibility of recruiting a larger sample. %This extended duration and the demanding nature of the tasks likely restricted the number of individuals willing and able to participate, thereby impacting the overall sample size. 
}

%is comparable with \hl{other studies}~\cite{}, the first replication includes a more limited number of subjects
This decreases the statistical power of our analysis, leading to the possibility of Type II error, i.e., failing to reject the NULL hypothesis when this is false.  %However, (1)~the Type I error is not affected, and thus, we can be confident in the results associated with the rejected NULL hypotheses, and (2)~the alignment between the results of the different experiments mitigates the problem associated with the reduced statistical power in the first replication. 
\textcolor{black}{A power analysis was not performed beforehand, as we focused on collecting the highest possible number of suitable participants. However, a power analysis was conducted after the experiments, to check the conclusion validity of our results. We performed a power analysis of both the paired $t$-tests (RQ1--RQ2) and the correlation tests (RQ3--RQ5), considering a medium effect size. For the $t$-tests, this resulted in a statistical power of $\sim 60\%$ for the original experiment and the second replication, and $\sim 25\%$ for the first replication. For the correlation tests, the statistical power is $\sim 66\%$ for the original experiment and the second replication, and $\sim 26\%$ for the first replication. These values are limited when considered individually---especially for the first replication. However, this is a family of experiments with a total of 50 participants. 
While individual experiments may suffer from limited power, 
the consistency of effects across multiple replications strengthens the credibility of the observed relationships. 
By aggregating the findings across experiments, the overall statistical power increases, reducing the likelihood of Type~II errors. In addition, the Type~I error is not affected, and thus, we can be confident in the results associated with the rejected NULL hypotheses. 
}

\section{Related Work}\label{sect:related}
\textbf{Evaluation of Language Workbenches.} Neverlang has been analysed in the study conducted by Cazzola et al.~\cite{CF22}, which employed various metrics, including coupling and cohesion, to assess its ease of use in facilitating variability in the creation of families of DSLs. 
For comparative analyses of existing language workbenches, the annual Language Workbench Challenge was launched in 2011. Each year, participants are tasked with implementing a given DSL using their workbenches as a foundation for discussion and comparison. Erdweg et al.~\cite{erdweg2013state,Erdweg15} present and discuss the results of these challenges up to the year 2015.
Kelly~\cite{kelly2013empirical} provides an overview of empirical research on language workbench comparisons, proposing an experimental design for future comparisons based on five criteria: feature coverage, lines of code, user satisfaction, time, and cost.
We are not aware of studies that evaluated language workbenches in terms of meta-language comprehensibility and acceptance.

\medskip
\noindent
\textbf{Evaluation of Languages.} Neverlang's meta-language can itself be regarded as a DSL, considering the ``programming languages'' field as the domain of interest. Therefore, we discuss related work on DSL evaluation. In this area, program comprehensibility has been evaluated by comparing the performance of users carrying out a series of tasks with one or more DSLs and a general purpose language (GPL). 

In~\cite{kosar2009influence,kosar2010comparing,kosar2012program}, Kosar et al.\ compared users' comprehension across different DSLs and GPLs.
The comprehension questionnaire used in these studies comprised 11~questions, categorised into learning, understanding, and evolving questions. We adopted the same template for our family of experiments. 
In \cite{kosar2009influence}, both the comprehension questionnaire and the cognitive dimensions framework from Green et al.~\cite{green1996usability} were used. Cognitive dimensions, in this case, are properties of the language used to evaluate the usability of programming languages and notations, which help understand how different aspects of a language affect ease of use and comprehension. These dimensions were mapped to specific questions in the questionnaire. 

The results of these studies showed that participants performed significantly better in terms of correctness when working with DSLs compared to GPLs, suggesting that DSLs are overall more comprehensible. The findings indicated a higher success rate and shorter completion time with DSLs, demonstrating better effectiveness and efficiency~\cite{kosar2009influence,kosar2010comparing,kosar2012program}. Moreover, participants consistently rated the DSLs as simpler \cite{kosar2012program}. Regarding cognitive dimensions, the closeness of mapping (how well the language’s constructs match the problem domain), viscosity (the effort required to make changes), and error-proneness (the likelihood of making mistakes) were identified as key factors affecting DSL comprehension~\cite{kosar2009influence}.

A similar experiment conducted by Kieburtz et al. \cite{kieburtz1996software} aimed to evaluate program development by comparing the use of a DSL and a GPL.
%, specifically Ada
In this study, four developers were tasked with creating message translation and validation modules for military systems. The results revealed that participants using the DSL were able to solve nearly three times as many tasks as those using the GPL. Furthermore, participants demonstrated a higher level of correctness in their tasks when utilising the DSL.

All this work indicates that DSLs are promising tools that can speed up and enhance the correctness of programming-related tasks. Thus, a DSL like Neverlang should, in principle, be preferred to a GPL when defining new languages, highlighting the relevance of our focus. With respect to this previous work, we are the first to focus on the evaluation of the DSL (i.e., the meta-language) of a language workbench. Compared to~\cite{kosar2009influence}, we consider a different cognitive dimension, i.e., WM capacity. The other dimensions~\cite{green1996usability}, which are inherent to the language and do not depend on the cognitive capacity of its users, will be taken into account in future work.

\section{Conclusion and Future work}\label{sect:conclusion}
This study presents a comprehensive evaluation of the comprehensibility and acceptance of Neverlang, a language workbench designed for modular development of programming languages.

Through a family of three experiments involving students, we assessed the comprehensibility of Neverlang’s meta-language and its acceptance in terms of ease of use, usefulness, and intention to use, as well as their relationship. 
The results show that more than two-thirds of users achieved sufficient comprehension, particularly at the syntactic level, and that perceived usefulness and intention to use were generally high, although ease of use emerged as a critical area for improvement. Moreover, the intention to use the tool appears to be driven by both its perceived usefulness and ease of use, highlighting the importance of addressing both factors to support broader adoption.

This study opens several avenues for future research to enhance the comprehensibility and acceptance of Neverlang. One important direction is to conduct comparative studies that evaluate Neverlang against other language workbenches. 
Another significant area involves extending the analysis to include participants from industry and academics with higher expertise and interest in language workbenches and language comprehension. 
Longitudinal studies are also essential to assess the long-term usability and reusability of Neverlang modules. Tracking users over an extended period will allow researchers to evaluate how well they can recognise and utilise reusable modules, providing insights into the tool’s efficiency and effectiveness in supporting modular and declarative programming. 
Additionally, a qualitative study aimed at identifying specific usability challenges and pain points could offer deeper insights into areas for improvement and enhance the overall user experience.
Another key direction for improvement is the enhancement of Neverlang’s ecosystem, including better documentation, an expanded set of libraries, and the development of multiple user interfaces—such as GUIs—tailored to different user profiles to improve accessibility and adoption.

%\IEEEtriggeratref{18}
%\bibliographystyle{IEEEtran}
%\bibliography{biblio}

\begin{acks}
Supported by the Italian MUR--PRIN 2020TL3X8X project T-LADIES\@: Typeful Language Adaptation for Dynamic, Interacting and Evolving Systems.
\end{acks}

\bibliographystyle{ACM-Reference-Format}
\bibliography{biblio_new}
%\bibliography{strings,biblio,dsl,my_work,pattern,splc}


\end{document}